\documentclass[twocolumn,showpacs,preprintnumbers,superscriptaddress,amsmath,amssymb,pra]{revtex4}

\usepackage{graphicx}
\usepackage{dcolumn}
\usepackage{bm}

\graphicspath{{./FIGURES/}}

\newcommand{\wc}{\omega_{\rm c}}

\newcommand{\kc}{\kappa_{\rm m}}

\newcommand{\kl}{\kappa_{\rm loss}}

\newcommand{\ka}{\kappa_{\rm a}}

\newcommand{\nt}{\bar{n}_{T}}

\newcommand{\na}{{n}_{{\rm{\mathcal{A}}}}}

\newcommand{\nac}{\bar{n}_{\mathrm{a}}}

\newcommand{\namp}{n_{\mathrm{amp}}}

\newcommand{\Var}{\mathrm{Var}}

\newcommand{\ainone}{\hat{a}_{\mathrm{in,1}}}
\newcommand{\aindone}{\hat{a}^{\dagger}_{\mathrm{in,1}}}
\newcommand{\aintwo}{\hat{a}_{\mathrm{in,2}}}

\newcommand{\ainj}{\hat{a}_{\mathrm{in},j}}

\newcommand{\aink}{\hat{a}_{\mathrm{in},k}}
\newcommand{\aindk}{\hat{a}^{\dagger}_{\mathrm{in},k}}

\newcommand{\binone}{\hat{b}_{\mathrm{in,1}}}
\newcommand{\bindone}{\hat{b}^{\dagger}_{\mathrm{in,1}}}
\newcommand{\bintwo}{\hat{b}_{\mathrm{in,2}}}
\newcommand{\bindtwo}{\hat{b}^{\dagger}_{\mathrm{in,2}}}

\newcommand{\aoutone}{\hat{a}_{\mathrm{out,1}}}
\newcommand{\aoutdone}{\hat{a}^{\dagger}_{\mathrm{out,1}}}
\newcommand{\aouttwo}{\hat{a}_{\mathrm{out,2}}}

\newcommand{\aoutj}{\hat{a}_{\mathrm{out},j}}

\newcommand{\aoutk}{\hat{a}_{\mathrm{out},k}}

\newcommand{\boutone}{\hat{b}_{\mathrm{out},1}}

\newcommand{\bouttwo}{\hat{b}_{\mathrm{out,2}}}

\newcommand{\A}{\hat{a}}
\newcommand{\Ad}{\hat{a}^{\dagger}}

\newcommand{\X}{\hat{X}}

\newcommand{\Y}{\hat{Y}}

\begin{document}

\preprint{}

\title{Accelerating dark-matter axion searches with quantum measurement technology}

\author{Huaixiu Zheng}
\affiliation{\textit{Department of Physics, Yale University, New Haven, Connecticut 06520-8120, USA}}

\author{Matti Silveri}
\affiliation{\textit{Department of Physics, Yale University, New Haven, Connecticut 06520-8120, USA}}

\author{R. T. Brierley}
\affiliation{\textit{Department of Physics, Yale University, New Haven, Connecticut 06520-8120, USA}}

\author{S. M. Girvin}
\affiliation{\textit{Department of Physics, Yale University, New Haven, Connecticut 06520-8120, USA}}

\author{K. W. Lehnert}
\affiliation{
JILA, University of Colorado and National Institute of Standards and Technology\\
Boulder, CO 80309-0440}

\date{\today}

\begin{abstract}
The axion particle, a consequence of an elegant hypothesis that resolves the strong-CP problem of quantum chromodynamics, is
a plausible origin for cosmological dark matter. In searches for axionic
dark matter that detect the conversion of axions to microwave photons, the quantum noise associated
with microwave vacuum fluctuations will soon limit the rate at which parameter space is
searched. Here we show that this noise can be partially overcome either by squeezing the
quantum vacuum using recently developed Josephson parametric devices, or by using superconducting qubits to count microwave photons.  The recently demonstrated ability of superconducting qubits to make QND measurements of microwave photons offers great advantages over destructive photon counting methods such as those using Rydberg atoms.
\end{abstract}

\pacs{Valid PACS appear here}
\maketitle

\section{\label{sec:Intro}Introduction and Organization}
The nature and origin of cosmological dark matter is an enduring puzzle
of modern physics. Likewise, the charge-parity (CP)
symmetry in the strong nuclear force seems implausibly well conserved. The hypothesis
of Peccei and Quinn resolves this ``strong CP problem'' by positing
a scalar field that couples to quarks and undergoes a spontaneous symmetry-breaking phase transition \cite{PecceiPRL77}.
Excitations of this field in its low energy phase, known as axions \cite{WeinbergPRL78, WilczekPRL78}, would
have the appropriate properties to act as a source of dark matter \cite{Abbott1983133, Dine1981, PRESKILL1983127}.

The rate of new experimental results that impact physics beyond the standard model is accelerating, with recent results from searches for dark matter particles \cite{PhysRevLett.111.251301, PhysRevLett.112.091303}, measurement of the electron dipole moment  \cite{ACMECollaboration17012014}, and detection of the Higgs boson \cite{Aad20121, ChatrchyanPLB12}. On balance, these recent results have strengthened the case for axionic dark matter. Consequently, it is important to test more rapidly the Peccei and Quinn hypothesis, particularly as it pertains to axionic dark matter \cite{BudkerPRX14,GrahamARNPS15}.

Searches for axionic dark matter exploit the coupling between the axion field $\mathcal{A}$ and
electromagnetism. Specifically, the axion field is believed to couple to the psuedo-scalar combination of electric field $\vec{E}$ and magnetic field $\vec{B}$ with a Lagrange density $g_{\gamma \gamma \mathcal{A}}\vec{E}\cdot\vec{B}\mathcal{A}$. The coupling constant $g_{\gamma \gamma \mathcal{A}}$ is proportional to the axion rest mass energy $m_a c^2$, tightly constraining the unknown parameters in the theory. In the presence of a static magnetic field, the axion field and electric field mix so that an axion may resonantly convert to a photon whose energy is approximately the axion rest-mass energy. Expressed as a frequency, the favorable energy range for such axion-derived photons is approximately 500~MHz to 500~GHz \cite{BradleyRMP03}.

\begin{figure}[ht]
  \includegraphics{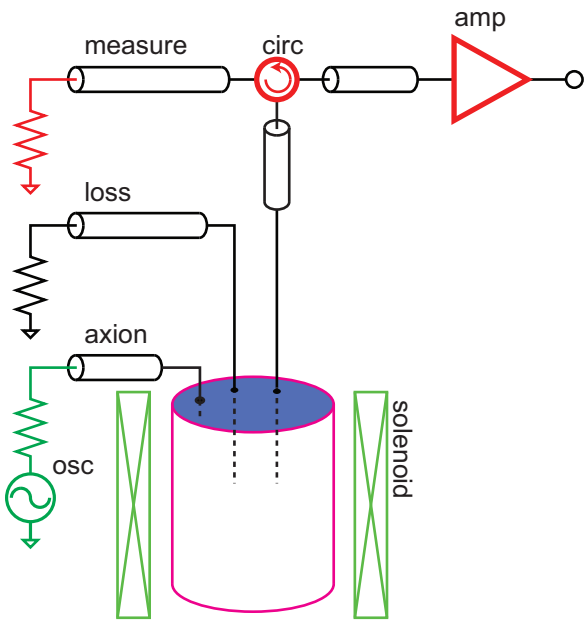}
  \caption{(Color online) Diagram of an axion dark matter search apparatus. A tunable microwave cavity is cooled far below ambient temperature and embedded in large magnetic field (solenoid), which---in the presence of an axion field---generates a feeble electric field oscillating a frequency $m_a c^2/\hbar$ modeled as if it were caused by a microwave oscillator (osc) with a large amplitude but weak coupling. The fluctuation and dissipation of the cavity are modeled as arising from three coaxial cables that protrude into the cavity mode. These ports extract energy and deliver noisy fields from the quantum-Nyquist noise of the resistors that terminate the cables. The ports are: a measurement port (measure), an internal dissipation port (loss) and, a possible interaction with the axion field (axion). The measurement port is in fact a coaxial cable whose coupling $\kc$ can be adjusted while the cavity is cold. It is distinct because the source of its fluctuations are experimentally accessible and its incident fields can be separated from its outgoing fields using a microwave circulator (circ). The outgoing fields are then measured by an amplifier (amp) to infer the cavity quadratures $\X_\mathrm{c}$ and $\Y_\mathrm{c}$.}\label{Fig:QuantumOpticsDiag}
\end{figure}

One type of axion dark matter search \cite{SikiviePRL83} uses a cryogenic and mechanically tunable microwave cavity to enhance sensitivity
to axion-derived photons (Fig.~\ref{Fig:QuantumOpticsDiag}). Specifically, if the
axion-generated electric field oscillates at the cavity's resonance frequency, the axion to photon conversion rate is enhanced by the cavity quality factor. By adjusting the cavity's resonance frequency, the range of favorable frequencies can then be scanned in a step-wise manner,
tuning the cavity to a new resonance frequency and waiting to average the cavity's thermal noise sufficiently well to resolve
the presence of any excess microwave photons caused by the coupling to the axion field. To reduce the background number of thermal photons that obscure the axion signal, the cavity is cooled well below ambient temperature.

Even if the cavity temperature were cold enough to completely freeze out this thermal background, existing
axion searches that use phase preserving amplifiers will be limited in their scan rate by a background noise associated with the quantum fluctuations
of the microwave vacuum. Due to the feeble axion-photon coupling, the mean number of cavity photons arising from axion conversion is very small ($\nac\sim 10^{-3}-10^{-6}$) \cite{BradleyRMP03}, while the quantum noise fluctuates with a variance of one cavity photon. The small size of this signal relative to the background quantum noise of $n_\mathrm{Q}=1$ photon poses a major challenge for the experimental detection of axions, as an axion signal is resolvable above the quantum noise background only after averaging $n_\mathrm{Q}/\nac^2=10^6-10^{12}$ independent realizations of the cavity state \cite{Dicke1946}. Even in a favorable region of the 500~MHz to 500~GHz frequency range, the time to scan just one octave of this range with a quantum limited measurement can easily exceed a year \cite{BradleyRMP03,BurbakerMISC15}. However, the quantum fluctuations are not a fundamental, unavoidable limit to axion detection; rather, they are a consequence of the fact that current experiments use phase-preserving linear amplifiers to measure the axion cavity field. Such linear, phase-preserving measurements fluctuate with a variance of at least one cavity photon essentially because they measure quantities with unfavorable commutation relations \cite{CavesPRD82, ClerkRMP10}.



%

In detail,
the harmonic oscillator Hamiltonian of the cavity mode of interest
$\hat{H}=\hbar\wc(\Ad_\mathrm{c} \A_\mathrm{c} + 1/2)$ can be written as $\hbar\wc(\X_\mathrm{c}^2+\Y_\mathrm{c}^2)/2$, where $\wc$ is the mode frequency,
{$\A_\mathrm{c}$ annihilates a cavity photon, $\X_\mathrm{c}=(\A_\mathrm{c} + \Ad_\mathrm{c})/\sqrt{2}$ and $\Y_\mathrm{c}= (\A_\mathrm{c} - \Ad_\mathrm{c})/(\sqrt{2}i)$ } are the cavity field quadratures. From the linear phase-preserving microwave amplifiers sensing the microwave field exiting the cavity, one can infer  $\X_\mathrm{c}$ \textit{and} $\Y_\mathrm{c}$.
Unlike a measurement of photon number $\Ad_\mathrm{c} \A_\mathrm{c}$, repeated measurements of $\X_\mathrm{c}$ and $\Y_\mathrm{c}$ fluctuate because $\X_\mathrm{c}$ and $\Y_\mathrm{c}$ do not commute with each other.  In addition---because the cavity field is inferred by measuring fields exiting the cavity---the cavity field itself must continuously decay, typically to a thermal state. Even if the cavity were to decay to its zero-entropy ground state, repeated measurements $\X_\mathrm{c}$ or $\Y_\mathrm{c}$ would fluctuate as the cavity would be in an energy eigenstate, which is not an eigenstate of either operator. For a cavity in equilibrium with an environment at temperature $T$ and measured with
a quantum limited amplifier, observations of $\X_\mathrm{c}$ and $\Y_\mathrm{c}$ will fluctuate with a variance $\Var(\X_\mathrm{c})+\Var(\Y_\mathrm{c})= \nt+1$,
where $\nt=[\exp(\hbar\omega_c/k_BT)-1]^{-1}$. The residual unit (photon) of noise present if $T=0$ is
the quantum noise, with half attributed to the vacuum fluctuations of the cavity mode itself and half attributed to the quantum-limited amplifier added-noise $\namp = 1/2$ arising from simultaneous measurement of canonically conjugate observables ($[\X_\mathrm{c},\Y_\mathrm{c}]=i$). The amplifier added-noise can by avoided by using a phase-sensitive amplifier \cite{CavesPRD82}, which noiselessly measures just one quadrature of the field exiting the cavity. But by itself, a phase sensitive amplifier confers no benefit in searching for an axion signal because it has not reduced the vacuum fluctuations and the factor two reduction in total measurement noise is compensated by the fact that on average half of the axion signal power will be in the unmeasured quadrature.

In this article, we describe two strategies to overcome the quantum noise that do increase the axion search rate. In the first strategy, we show how quantum squeezing of microwave fields can be used to accelerate the axion search in proportion to the amount of achievable squeezing. Furthermore, we show that the squeezing apparatus can be used to evade thermal noise
just as effectively as it evades quantum noise. Finally, we argue that existing quantum microwave measurement technologies, namely Josephson parametric devices, are already sufficiently well developed that this concept can be deployed in current axion searches.

The second strategy involves use of superconducting qubits to count individual photons.  The fact that photon counting methods can beat the standard quantum limit of linear phase-preserving amplifiers has been known for some time, and a number of axion detection proposals and experimental efforts have been made based on using ionization of Rydberg atoms to detect individual microwave photons \cite{MATSUKI1991523,Ogawa_PhysRevD.53.R1740,CARRACK-II-TADA1999164,SinglePhotonRydberg_Tada2006488,Rydberg2008,BradleyRMP03}.  Based on the tremendous experimental progress in superconducting qubits over the last 15 years, it is now possible to routinely detect single microwave photons with high efficiency.  Essential to our proposal is the fact that this detection can be made highly quantum non-demolition (QND) \cite{JohnsonNatPhys10,SunNat14}. This offers an enormous advantage over Rydberg atom methods because the QND nature allows the measurement to be repeated hundreds of times.  Optimal Bayesian filtering of the measurement record enhances the quantumn efficiency and strongly suppresses the rate of false positives (dark counts).
With further development of superconducting qubits operated as photon counters, this concept holds the greatest potential for increase in axion search rates.

\section{\label{sec:QntSqzApp} Quantum squeezing approach}

Although repeated measurements of $\X_\mathrm{c}$ and $\Y_\mathrm{c}$ must fluctuate, quantum squeezing can be used to avoid this quantum noise when attempting to infer the presence of an axion field through its displacement of $\X_\mathrm{c}$ and $\Y_\mathrm{c}$. In an idealized example, the cavity could first be prepared in a pure squeezed state with $\Var(\X_\mathrm{c})\ll \Var(\Y_\mathrm{c})$ and saturating the Heisenberg uncertainty $\Var(\X_\mathrm{c})\Var(\Y_\mathrm{c})=1/4$. The evolution of this state under the cavity Hamiltonian ensures that $\X_\mathrm{c}$ recovers its minimum variance periodically with period $\pi/\wc$. A subsequent (noiseless) measurement of \textit{only} $\X_\mathrm{c}$ at an instant of minimum variance would resolve an arbitrarily small (axion induced) displacement of $\X_\mathrm{c}$ in the limit of arbitrarily large squeezing \cite{Yurke1984}.
Succinctly, in the limit of large squeezing one prepares the cavity in an eigenstate of the measured quantity, thus inferring one quadrature of the axion field without quantum fluctuations, but forgoing knowledge of the other quadrature and thus reducing the detectable power by half. This concept can be extended to measure both quadratures of the axion field by introducing a second cavity (not coupled to the axion field) with quadratures $\X_2$ and $\Y_2$. The EPR-like observables $\hat{Q}=\X_\mathrm{c}+\X_2$ and $\hat{P}=\Y_\mathrm{c}-\Y_2$ commute and therefore the two-cavity system can be prepared in a simultaneous eigenstate of both $\hat{Q}$ and $\hat{P}$. Just as for the single-mode squeezing concept, if one can both prepare the appropriate two-cavity eigenstate and arrange to measure only $\hat{Q}$ and $\hat{P}$ both axion quadratures can be measured noiselessly \cite{YurkePRA86, TsangPRX12}.

\subsection{\label{sec:QntOptMod} Quantum Optics Model}

To understand how this idealized notion can be implemented in practice, we express the cavity axion search experiment in the formalism of quantum optics.
The cavity mode exchanges energy with three distinct environments (ports): a measurement port engineered to couple the cavity to an amplifier through a transmission line, a port modeling the cavity's internal loss, and a port associated with the putative axion-photon interaction  (Fig.~\ref{Fig:QuantumOpticsDiag}).  We first transform to a frame rotating at the cavity's resonance frequency defining {$\A_\mathrm{c}(t)\rightarrow \A_\mathrm{c}(t)\exp(-i \wc t)$} and,
with this transformation, the Heisenberg-Langevin equation of motion for the cavity field is simply
\begin{equation}
  \frac{d\A_\mathrm{c}}{dt} =  - \frac{\kappa}{2} \A_\mathrm{c}(t) + \sum_j\sqrt{\kappa_j}\ainj \label{eq:CavEOM}
\end{equation}
where $\ainj \in \{\hat{a}_{\mathrm{in,m}},\hat{a}_{\mathrm{in,loss}},\hat{a}_{\mathrm{in,a}}\}$ are
the annihilation operators (in the same rotating frame) of the modes of the environment with commutation relations $[\ainj(t), \aindk(t')]=\delta(t-t')\delta_{jk}$. They model the input fields incident on the measurement, loss, and axion port respectively. Likewise $\kappa_j \in \{\kc, \kl, \ka\}$
are the rates that the cavity energy decays to the three ports, and $\kappa=\sum_j \kappa_j$.  The field exiting a port is related to the incident field at that port and to the cavity mode according to input-output relations\cite{ClerkRMP10} as
\begin{equation}
  \aoutj = \ainj  - \sqrt{\kappa_j}\A_\mathrm{c}(t). \label{eq:OutField}
\end{equation}
These simple linear equations of motion and input-output relations can be solved in the Fourier domain as $\aoutj(\omega) = \sum_{k}\chi_{jk}\aink(\omega)$, where
\begin{equation}
\chi_{jk}(\omega)=\frac{-\sqrt{\kappa_j}\sqrt{\kappa_k} + \left(\kappa/2 + i \omega\right)\delta_{jk}}{\left(\kappa/2 + i \omega\right)} \label{eq:AxCavSusceptMat}
\end{equation}
is a $3\times3$ susceptibility matrix.

These susceptibilities and the noise of the input fields determine the noise in the output fields. Because the rate at which axions
convert to photons is so much slower than the other dissipative rates ($\ka \ll \kl \sim \kc$),
we can consider the noise properties in the absence of the axion port (See Appendix~\ref{sec:App:FullTMSwithLoss}).
Assuming that the fields incident at the loss and measurement ports
are in equilibrium with the same thermal environment at temperature $T$, the noise of the incident fields are characterized by a covariance matrix  with elements $\langle[\ainj(\omega')]^\dag \aink(\omega)\rangle=2\pi\nt \delta(\omega-\omega')\delta_{jk}$ and $\langle\ainj(\omega) [\aink(\omega')]^\dag \rangle=2\pi(\nt+1) \delta(\omega-\omega')\delta_{jk}$ . As the two environments are at the same temperature, the noise of the outgoing fields is also $\langle[\aoutj(\omega')]^\dag \aoutk(\omega)\rangle=2\pi\nt \delta(\omega-\omega')\delta_{jk}$ and $\langle\aoutj(\omega) [\aoutk(\omega)]^{\dag}\rangle=2\pi(\nt+1) \delta(\omega-\omega')\delta_{jk}$ \cite{ClerkRMP10}.

Although the output noise is frequency independent, an axion signal must still have a frequency close to the cavity's resonance to be detectable. The susceptibility of the field exiting the measurement port to an incident axion field $|\chi_\mathrm{ma}|=|\sqrt{\kc \ka}/(\kappa/2 +i\omega)|$ is maximized at the cavity's resonance frequency $\omega=0$. Consequently, the ratio $\alpha(\omega)$ of the axion-induced signal power to the total noise power will be proportional to $|\chi_\mathrm{ma}(\omega)|^2$ and similarly maximized at cavity resonance.

\begin{figure}[ht]
  \includegraphics[width=3.4 in]{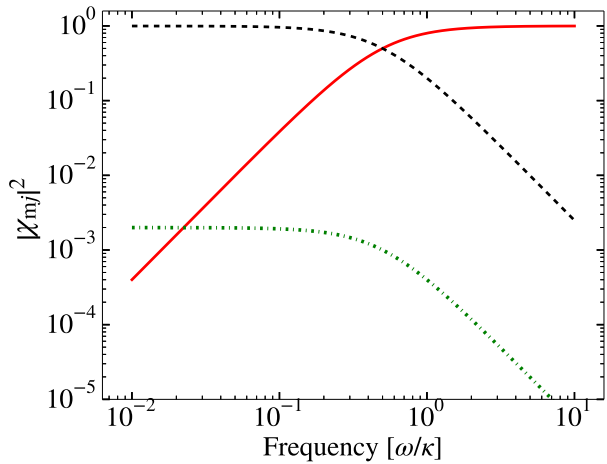}\\
  \caption{(Color online) Susceptibility matrix: The square magnitude of susceptibility matrix elements: $|\chi_\mathrm{mm}|^2$ (red solid), $|\chi_\mathrm{ml}|^2$ (black dashed), $|\chi_\mathrm{ma}|^2$ (green dashed dotted) are plotted as a function of Fourier frequency ($\omega$) detuned from the cavity's resonance, with $\ka=\kappa/1000$ and $\kc=\kl$.  Even this value of $\ka\ll\kappa$ is implausibly large, but chosen so that all of the elements can be plotted on the same logarithmic scale.}\label{Fig:SusceptibilityFigure}
\end{figure}

The wasteful aspect of the cavity measurement becomes evident when optimizing $\alpha(\omega)$ over $\kappa \approx \kl+\kc$. The axion power converted to microwave power diminishes with increasing $\kappa$. Because the axion cavity must reside in a large magnetic field, it is made from a non-superconducting metal and $\kl$ cannot be reduced arbitrarily \cite{BradleyRMP03}. Fixing $\kl$ at the smallest achievable value and maximizing $\alpha(\omega)$ over $\kc$ reveals $\alpha(\omega=0)$ is maximum for $\kc=\kl$ where the noise exiting the cavity measurement port comes entirely from noise incident at the loss port ($\chi_\mathrm{ml}(0)=1$). This on-resonance and critically coupled sensitivity is the technically-limited axion sensitivity; it can only be improved by technical advances that reduce $\kl$ or $\nt$, not by quantum squeezing. (The quantum states of the degrees of freedom of the loss port are by definition inaccessible.) If the loss port were the only source of noise, an axion search would maintain this technically-limited sensitivity at frequencies increasingly far from cavity resonance because $\chi_\mathrm{ma} (\omega)$ and $\chi_\mathrm{ml}(\omega)$ diminish together, remaining in proportion. But away from cavity resonance the susceptibility of the measurement-port output to the measurement-port input ($\chi_\mathrm{mm}(\omega)$) grows and $\alpha(\omega)$ decreases (Fig.~\ref{Fig:SusceptibilityFigure}) due to the noise incident at the measurement port. Consequently, if one just surpasses the requisite sensitivity to detect an axion signal on resonance, the range of frequencies over which one might detect an axion signal is limited to $\kappa$. For typical parameters, the spectral width of the axion line is between $\kappa/10$ and $\kappa/100$ thereby testing 10 to 100 independent values of the axion mass simultaneously \cite{AsztalosPRD04, BurbakerMISC15}. On the other hand, if one could reduce the noise entering through the measurement port by squeezing it by an arbitrarily large amount, the sensitivity achieved at cavity resonance could be extended over a frequency range arbitrarily wider than $\kappa$ thus searching a wider range of axion masses simultaneously.
\subsection{Ideal analysis of two-mode squeezing}\label{sec:TwoModeSqueezer}

In effect a sequence of two, two-mode squeezing operations accomplishes this reduction in noise \cite{Yurke1984, Flurin2012}. A two-mode squeezer (TMS) transforms a pair of input modes as
\begin{eqnarray}\label{eq:TMSeqs}
\boutone&=&\binone \sqrt{G} + \bindtwo \exp(i\phi)\sqrt{G-1}\\
\bouttwo&=&\bintwo \sqrt{G} + \bindone \exp(i\phi)\sqrt{G-1},
\end{eqnarray}
where $G$ is the power gain (related to the squeezing parameter $r$ as $\sqrt{G}=\cosh(r)$), and $\phi$ is the phase of the pump field that drives the squeezing process. Considering only the noise properties of $\boutone$ and ignoring $\bouttwo$, a two-mode squeezer is simply a phase preserving amplifier with power gain $G$ and added noise determined by the fluctuations of the field $\bintwo$. With $\bintwo$ in a thermal state at temperature $T$, the amplifier added noise is $\nt + (1/2)$.  If both outputs are considered, an ideal two-mode squeezer performs a unitary, noiseless transformation on the two input fields. With both input modes in a vacuum state, the output is a pure two-mode squeezed and entangled state with strong correlations between the two modes\cite{CavesPRD82, Yurke1984}.

\begin{figure}[ht]
  \includegraphics[width=3.4 in]{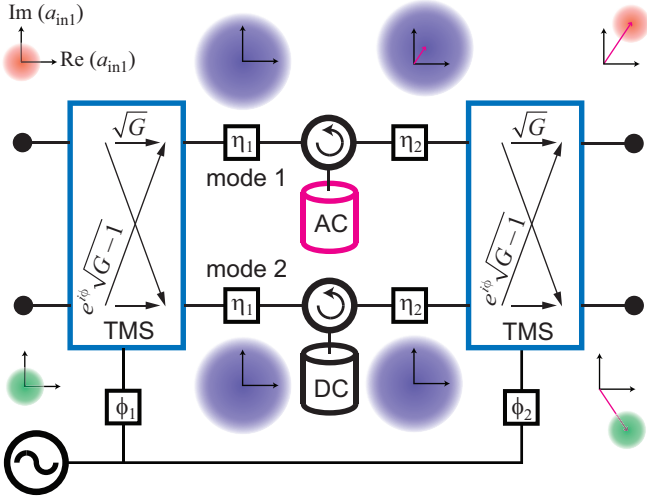}\\
  \caption{(Color online) A schematic illustration of the benefit of squeezing. The circular regions plotted on the axes (e.g. $\mathrm{Re}(\ainone),\ \mathrm{Im}(\ainone)$) depict the transformation of two vacuum states as they pass through the axion receiver. The two modes are amplified and entangled at the first two-mode squeezer (TMS). One mode is then injected into the axion cavity (AC), potentially acquiring a displacement (greatly exaggerated) associated with an axion field, while the other mode is simply injected into a delay cavity (DC) matched to the susceptibility of the axion cavity. The outputs of these two cavities are directed to a second two-mode squeezer with gain and phase chosen to undo the operation of the first squeezer. Transmission loss that may be present between the cavities and the squeezer elements is modeled with beam splitter elements with transmission $\eta_1$ or $\eta_2$.}\label{Fig:ConceptFigure}
\end{figure}

Figure~\ref{Fig:ConceptFigure} illustrates the scheme for using these strong correlations to circumvent quantum noise in the axion search. Two microwave modes ($\ainone$ and $\aintwo$), propagating in transmission lines, are the inputs to the first TMS with power gain $G_1$ and pump phase $\phi_{1}$. Mode~1 is then injected into the axion cavity while mode~2 is injected into a second resonant cavity (delay cavity) designed to match the susceptibility of the axion cavity by designing its measurement port coupling rate  $\kappa_{\mathrm{dm}}=\kc$ and internal loss rate $\kappa_{\mathrm{dl}}=\kl$ to equal those of axion cavity. (See appendix~\ref{sec:App:FullTMSwithLoss} for a discussion of this choice.) The outputs from these two cavities are then fed to a second TMS with gain $G_2=G_1$ and pump phase $\phi_{2}=\phi_{1}-\pi$ chosen to undo the first squeezing operation thus recovering the initial input modes at the output of the second TMS ($\aoutone=\ainone$ and $\aouttwo=\aintwo$). If however mode~1 is displaced in phase space by an axion field, this displaced component appears at the output modes $\aoutone$ and $\aouttwo$,  but amplified by the gain of the second TMS. In principle, one mutually orthogonal quadrature of each output mode could then be measured noiselessly with two phase-sensitive amplifiers. This arrangement amplifies both quadratures of an axion induced microwave signal but without adding noise and, more surprisingly, without amplifying the vacuum noise! More practically, the same sensitivity can be reached by operating the second TMS such that $G_2 \gg G_1$, and measuring both quadratures of $\aoutone$. With sufficient gain, any noise added after the second TMS is negligible.  In the limit of large $G_1$ and $G_2$, the scheme realizes the concept of Sec.~\ref{sec:QntSqzApp}; the first TMS prepares the two-cavity system in a joint eigenstate of $\hat{Q}$ and $\hat{P}$ and the second TMS measures $\hat{Q}$ and $\hat{P}$ but without measuring $\X_\mathrm{c}$ and $\Y_\mathrm{c}$ separately.


\subsection{Search Rate Enhancement}\label{sec:SerRtEnh}


To understand the relationship between the bandwidth increase and squeezing we calculate $\alpha(\omega)$ in a system like Fig.~\ref{Fig:ConceptFigure}, but accounting for various sources of loss. We find a susceptibility matrix $\mathbf{\Xi}$ for the two TMS concept including transmission loss in both modes between the first TMS and the two cavities and between the cavities and the second TMS. In addition, we include the loss within both cavities. Because it describes the scattering from 9 input modes---$\ainone,\ \aintwo$, the axion cavity loss-port, the axion cavity axion-port, the delay cavity loss-port, and 4 input modes associated with transmission loss---to 9 output modes, $\mathbf{\Xi}$ is a simple to calculate but unwieldy to represent $18 \times 18$ matrix. (Its construction is detailed in appendix~\ref{sec:App:FullTMSwithLoss}).

To express this transformation compactly, we introduce two 18-element vectors, one describing the nine input fields $\vec{u}=[\ainone,...,\hat{a}_{\mathrm{in},9},\hat{a}_{\mathrm{in},9}^{\dagger},...,\aindone]^{\mathrm{T}}$ and one describing the output fields $\vec{y}=[\aoutone,...,\hat{a}_{\mathrm{out},9},\hat{a}_{\mathrm{out},9}^{\dagger},...,\aoutdone]^{\mathrm{T}}$, such that $\vec{y}(\omega)=\mathbf{\Xi}(\omega) \vec{u}(\omega)$. The power spectral density at the output of mode~1 due to an axion signal $\mathbf{S}_\mathrm{out1,ax}=([\mathbf{\Xi}]_{1,4}^2(\na) + [\mathbf{\Xi}]_{18,15}^2(\na+1))/2$ can be found from the elements of this matrix as
\begin{equation}\label{eq:OutputAxionSuscept}
\mathbf{S}_\mathrm{out1,ax} = \frac{G_2  \ka \kc(\na+(1/2))}{(\kappa/2)^2 + \omega^2 },
\end{equation}
where $\na(\omega)$ is the spectral-density of the \textit{incident} axion field.
The covariance matrix describing the total output noise is compactly written $\langle[\vec{y}(\omega')]^{\dagger}\vec{y}^{\mathrm{T}}(\omega)\rangle=2 \pi \mathbf{S}_{\mathrm{out}}\delta(\omega'-\omega)$, where $\mathbf{S}_{\mathrm{out}}$ is the output spectral density matrix and in the expression $[\vec{y}(\omega')]^{\dagger}$ the Hermitian conjugate does not transpose the vector. This matrix can be found as $\mathbf{S}_{\mathrm{out}} = \mathbf{\Xi}^{*}(\omega) \mathbf{S}_{\mathrm{in}} \mathbf{\Xi}^{\mathrm{T}}(\omega)$, where $\mathbf{S}_{\mathrm{in}} = \mathrm{diag}[\nt,...,\na,...,\nt,\nt+1,...,\na+1,...,\nt+1]$ is the input noise spectral density matrix. (Again, for the purpose of determining the noise background upon which the tiny power spectral density $\mathbf{S}_\mathrm{out1,ax}$ must be detected, we calculate the total output noise assuming $\ka=0$.)

The total power spectral density at the output of mode~1 is $\mathbf{S}_\mathrm{out1,tot}=([\mathbf{S}_{\mathrm{out}}]_{1,1} + [\mathbf{S}_{\mathrm{out}}]_{18,18})/2$. We first consider  {the ratio of signal power to noise power spectral density,} $\alpha(\omega)$, in the absence of transmission loss and in the limit of large $G_2$ where we find
\begin{widetext}
\begin{equation}\label{eq:Ch1Variance}
\alpha(\omega) \equiv \frac{\mathbf{S}_\mathrm{out1,ax}}{\mathbf{S}_\mathrm{out1,tot}} = \frac{(\na + 1/2)}{(1+2\nt)}\frac{\ka \kc }{\beta(\omega)\left[(2G_1-1)+2\sqrt{G_1^2-G_1}\cos(\phi_{2}-\phi_{1}) + \kc \kl /\beta(\omega)\right]},
\end{equation}
\end{widetext}
and $\beta(\omega)=[(\kc - \kl)/2]^2+\omega^2$. This expression shows the expected behavior. If the two TMS's are operated such that $\phi_{2}-\phi_{1}=\pi$, $\alpha(\omega)$ is maximized. In the absence of squeezing ($G_1=1$) $\alpha(\omega)$ is maximum for $\kc=\kl$ and this same maximum can be achieved in the $G_1\rightarrow \infty$ limit, but for any
value of $\kc$. Thus by using squeezing and operating the cavity in the over coupled limit, the axion search can be accelerated.

\begin{figure}[ht]
  \centering
  \includegraphics[width = 3.4 in]{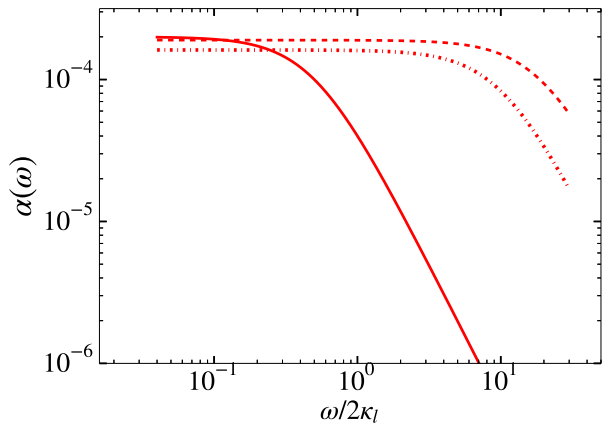}\\
  \caption{(Color online) Axion sensitivity. The plots show the estimated axion signal spectral density to noise spectral density ($\alpha$) as a function of frequency  {(relative to the cavity frequency)} with $\na \ka/\kl = 2\times10^{-4}$, for three different receivers. The axion receiver with a critically coupled cavity and without squeezing ($\kc=\kl$ and $G_1 = 1$, solid) has the highest sensitivity at resonance and a bandwidth of $\kappa$. The receivers with over-coupled cavities and with squeezing ($\kc=19\kl$ and $G_1 = 5$, dotted), ($\kc=19\kl$ and $G_1 = 20$, dashed) both have slightly lower sensitivity at cavity resonance but much larger bandwidths. In particular, the receiver with $G_1 = 20$ preserves its sensitivity over a 40 fold larger band.}\label{fig:SqueezBen}
\end{figure}

In Fig.~\ref{fig:SqueezBen} we plot $\alpha(\omega)$ comparing a critically coupled ($\kc=\kl$) axion receiver without squeezing to a receiver with ten-fold over coupled ($\kappa=10\times(2\kl)$) cavity bandwidth. To achieve the same sensitivity on resonance would require infinite squeezing, but 90\% of the critically coupled sensitivity can be achieved with $G_1=20$.  This small reduction in on resonance sensitivity is accompanied by a 40 fold increase in measurement bandwidth and consequently an almost 40 fold increase in scan rate.

The situation is not as favorable if there is transmission loss between the two TMSs. In effect, losing a fraction $1-\eta$  of microwave power  diminishes the amount of squeezing that can be delivered to the second TMS irrespective of $G_1$. Consequently there is a finite optimum value of $\kc$ and a finite increase in axion search rate for a given transmission loss. We show this result by modeling the loss as caused by a fictitious beam splitter, which directs a fraction $1-\eta$  of the power in mode~1 and 2 to unmeasured ports. In addition, noise enters mode~1 and 2 through these open ports. To quantify the benefit of squeezing in the presence of transmission loss, we find the rate at which an axion experiment can be scanned through frequencies. At each setting of the cavity frequency, one must wait for a time
\begin{equation}\label{eq:IntegrationTime}
\tau\approx(d_{\mathrm{SNR}}^2)/(\alpha^2(\omega=0) B_{w}),
\end{equation} where $B_{w}$ is the putative axion signal bandwidth, and $d_\mathrm{SNR}$ is the  {desired} threshold signal to noise ratio (SNR) for detecting an axion \cite{Dicke1946,KraussPRL85}. After averaging the noise power for time $\tau$ the cavity frequency can be tuned by the characteristic width of $\alpha^2(\omega).$ Thus one can scan frequency at an average rate $d \wc/dt \propto \int_{-\infty}^{\infty}d\omega\, \alpha^2(\omega)$. Even without squeezing, the axion scan rate is optimized not exactly at critical coupling but at the over-coupled value of $\kc=2\kl$ \cite{KraussPRL85}.

\begin{figure}[ht]
  \centering
  \includegraphics[width=0.49\textwidth]{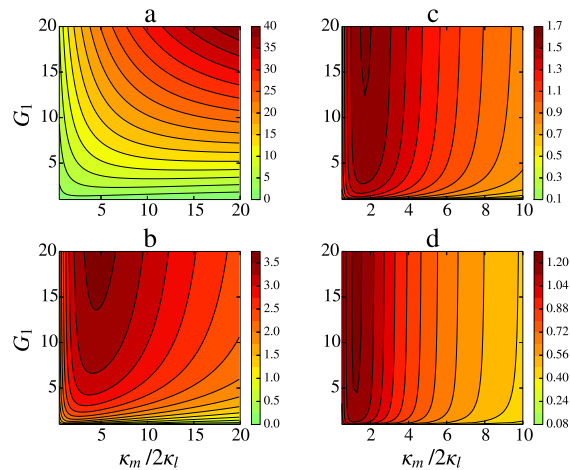}
  \caption{(Color online) Scan rate acceleration. The increase in the scan rate due to squeezing is shown as a function of $G_1$ and $\kc$ in the contour plots for a.) $\eta_1 = \eta_2 =1$, b.) $\eta_1 = \eta_2 =0.9$, c.) $\eta_1 = \eta_2 =0.7$, d.) $\eta_1 = \eta_2 =0.5$. In all plots $\phi_{1}=0$, $\phi_{2}=\pi$ and $G_2=1000$.}\label{fig:SqueezBenA}
\end{figure}

We find the ratio of the scan rate with squeezing ($G_1>1$) to the optimally over-coupled scan rate without squeezing ($G_1=1$, $\kc=2\kl$). Fig.~\ref{fig:SqueezBenA} plots this ratio as function of $G_1$ and $\kc$ for different values of loss $\eta_1$ between the first TMS and the cavities
and loss $\eta_2$ between the cavities and the second TMS. In the complete absence of loss the scan rate can be accelerated arbitrarily for arbitrary squeezing. For $\eta_1 = \eta_2 = 0.9$ one can almost quadruple the scan rate while for $\eta_1 = \eta_2 = 0.5$, the scan rate is just 25\% improved, highlighting the importance of low-loss microwave connections for quantum enhanced sensing.

Although we envision deploying squeezing to circumvent the quantum noise in an axion search, in the calculation of search rate acceleration we have not assumed that quantum noise is dominant ($\nt \ll (1/2)$). As long as all sources of noise are at the same temperature, the same speed up is realized when dominated by thermal noise, as suggested by the single factor of $1/(2\nt +1)$ in Eq.~\ref{eq:Ch1Variance}. Indeed, as the analysis is restricted to Gaussian states transformed by linear operations and measured with linear detectors, the quantum dynamics are quite hard to distinguish from classical dynamics \cite{BartlettPRL02}. For such situations, quantum mechanics is apparent only because the sources of thermal noise produce fluctuations even at zero temperature. In searches for axions or axion-like particles that extend to lower frequencies where one cannot achieve $\nt \ll (1/2)$, thermal squeezing is just as beneficial as quantum squeezing.

We conclude by noting that most of the technical challenges associated with implementing a quantum squeezed receiver within an axion search have already been addressed. In particular, Josephson parametric amplifiers (JPAs), which prepare and analyze squeezed states, are already being used in an axion search \cite{ShokairIJMPA14}. The need to tune the resonance frequency and loss of the delay cavity is no more complex than tuning the resonance frequency and coupling of the axion cavity, a routine task in an axion search \cite{AsztalosPRD04,AsztalosPRL10}. Finally, there has been substantial recent progress in improving microwave squeezing and amplification technology. Indeed there are recent demonstrations of both single-mode \cite{MalletPRL11} and two-mode squeezed states \cite{Flurin2012,KuPRA15} being prepared and analyzed by Josephson parametric devices. Currently, some single-mode devices have widely tunable amplification bands that can track the octave tuning range of an axion cavity \cite{CastellanosBeltranNPHYS08,MutusAPL13}. Although the two-mode devices have not yet shown such wide tunability, there does not seem to be a technical reason why they could not be adapted to be more tunable.

The real challenge lies in reducing transmission losses, as transmission efficiency in representative experiments can exceeded $\eta = 0.5$ \cite{MurchNAT13} but have not yet reached $\eta > 0.7$. Nevertheless, even a doubling of the axion search rate is well worth the effort because the quantum-limited time to search the 4 -- 8~GHz range may be several years.


\section{Photon Counting Approach}
In this section, we describe the second strategy we take to accelerate the search of axion particles: designing a single-photon detecter using superconducting qubits.
The development of quantun non-demolition measurements of microwave photon number using transmon qubits \cite{JohnsonNatPhys10,SunNat14}
makes it possible to perform high-fidelity readout of the cavity photon states.
We take advantage the fact that in circuit QED, measurements of photon number parity that are  nearly perfectly QND (99.8\%) have been demonstrated experimentally \cite{SunNat14}.  Such measurements can therefore be repeated hundreds of times  to improve the photon counting efficiency and dramatically suppress the dark count rate.
Key to this is processing the measurement data record with an optimal Bayesian smoothing algorithm.

\begin{figure}[tbh]
\centering
\includegraphics[width=0.49\textwidth]{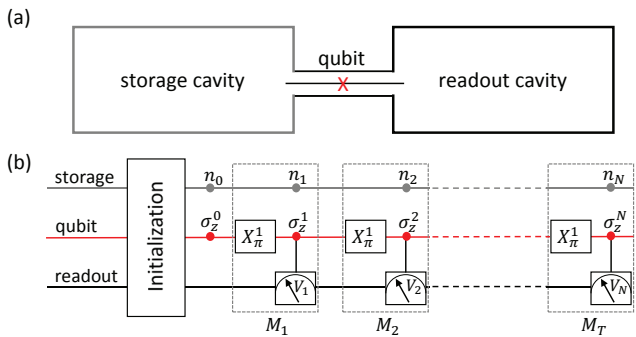}
\caption{(Color online) (color online) (a) Single photon detector using two cavities and one superconducting qubit.
(b) The protocol to measure the storage cavity photon number. After the initialization, the qubit state is $\sigma_{z}^{0}$ (g or e) and the cavity photon number is $n_{0}$
  ($0$ or $1$). In each measurement $M_{i}$, a one-photon selective $\pi$-pulse is implemented on the qubit, followed by a single shot dispersive measurement of the qubit state with a readout voltage $V_{i}$
 . Such measurement is repeated $N$
  times. }
\label{fig:Fig1-setup}
\end{figure}

\begin{figure*}[t!]
\centering
\includegraphics[width=0.9\textwidth]{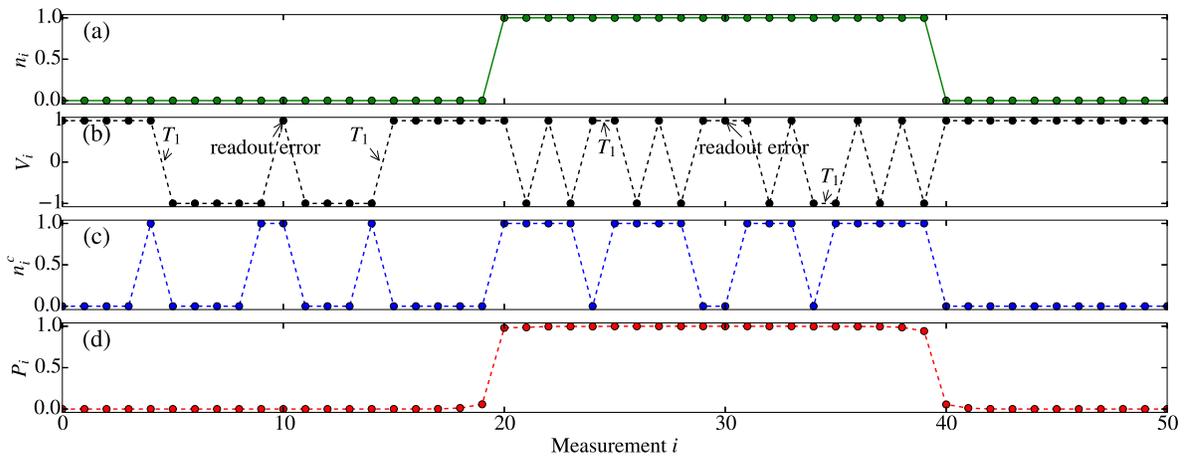}
\caption{(Color online) An example of (a) photon state $n_i$ and (b) the readout voltage $V_i$
after digitizing qubit $|g\rangle$ to $+1$ and $|e\rangle$ to $-1$.
$T_{1}$ jumps and readout errors are pointed out by the arrows. (c)
Photon number $n_i^{c}$ inferred from correlating the neighboring readout voltages $V_i$ and $V_{i+1}$.
(d) Probability $P_i=P(n_i=1|V_1\cdots V_N)$ that one photon is present in the cavity of the time of the $i^{\text{th}}$ measurement
obtained from applying the Bayesian smoothing algorithm to the entire measurement record [Eqs.\,(\ref{eq:CondP})-(\ref{eq:FB})]. The lines are a guide to eye.}
\label{fig:Fig2-example}
\end{figure*}

\subsection{Circuit-QED Setup and Protocol}

We propose to use a superconducting transmon qubit \cite{Koch2007,Schreier2008,PaikPRL11,RIgetti_PhysRevB.86.100506} coupled to two microwave cavities: one storage cavity and one readout cavity as in Ref.~\cite{CatStateVlastiakisScience2013}, see Fig.\  \ref{fig:Fig1-setup}(a).
In the presence of a strong static magnetic field, the storage cavity is used to capture photons resonantly converted from axions \cite{SikiviePRL83,SikiviePRD85}, and has a high quality factor $Q_{c}=\omega_c/\kappa$
for storage of photons. It can be either a copper cavity or a hybrid normal-superconducting cavity \cite{LamoreauxPRD13, XiPRL10}. The readout cavity is a standard three-dimensional superconducting cavity \cite{PaikPRL11} with a low quality factor
for fast readout of the qubit state \cite{CatStateVlastiakisScience2013}. The qubit and the cavities are detuned from each other so that they are in the dispersive coupling regime, which is described by the Hamiltonian
\begin{equation}
H/\hbar=(\omega_{q}-\Delta \hat{a}_c^{\dagger}\hat{a}_c)|e\rangle\langle e|+\omega_{c}\hat{a}_c^{\dagger}\hat{a}_c,
\end{equation}
where $\hat{a}_c^{\dagger}$ and $\hat{a}_c$ are the creation and
annihilation operators of the storage cavity, respectively and $|e\rangle$
is the excited state of the qubit. The dispersive coupling $\Delta$
gives rise to a cavity photon-number-dependent shift of the qubit
frequency. The qubit transition frequency $\omega_{q}$ and the cavity
resonant frequency $\omega_{c}$ are detuned such that $\Delta\gg\kappa$.
The readout cavity couples to the qubit in the same manner as the storage cavity and is used to perform QND measurements of the qubit state.

In this work, we consider only the zero-photon and one-photon states
since the mean photon occupation $\bar{n}$ in the cavity (from both
thermal and axion-derived photons) is expected to be much lower than
one.
We follow the circuit-QED procedures used in
the quantum non-demolition detection of single microwave photons \cite{JohnsonNatPhys10,Narla15}
and the photon number parity measurement.  Fig.\
\ref{fig:Fig1-setup}(b) shows the protocol. First, we apply a selective
$\pi$ pulse $\mathbf{\mathrm{X}}^1_{\pi}$ on the qubit which flips the qubit state if and only if there is exactly one photon in the cavity. Next, the qubit state $\sigma_{z}$ is
measured by the readout cavity using the qubit-state dependent cavity
response \cite{KirchmairNat13}. The above two steps give one
interrogation of the photon number in the storage cavity: the qubit
state is flipped if and only if there is one photon in the storage
cavity. By repeating the interrogation, we can monitor quantum jumps
of photons into and out of the storage cavity \cite{SunNat14}.  Importantly,  this can be done in a manner which is very nearly quantum non-demolition \cite{SunNat14}.

By correlating neighboring measurement results of
the qubit state, the cavity photon number can be inferred: if the qubit
state does not flip, then the cavity is in the zero-photon state; otherwise,
there is one photon in the cavity. However, in reality the above ideal
analysis suffers from several imperfections including errors in the readout of the qubit and the fact that the $\pi$ pulse may not be perfectly selective on photon number.  In addition there are $T_1$ decay processes in which the qubit decays from the excited to ground state or jumps up from ground to excited.  While it is relatively straightforward to thermally equilibrate the photons in the cavity using cold filtering, this is not true for transmon qubits.  For unknown reasons (possibly associated with superconducting quasiparticle production) they are often out of equilibrium and the steady state can have $p\sim10\%$ probability to be
in the excited state \cite{SunNat14}. Therefore, the qubit jump up process is included to account
for this excited state probability.
 {The rates $\Gamma_{\uparrow (\downarrow)}$ for these processes obey $\Gamma_\uparrow+\Gamma_\downarrow=1/T_1$ and $\Gamma_\uparrow = p/T_1$.}

Fig.\ \ref{fig:Fig2-example}(a)-(c) shows a model example of a series of qubit
measurements and the effect of imperfections on the correlation-inferred
photon number. Each $T_{1}$ jump (up or down) introduces a single
error in the photon number measurement. The readout error comes into play because
there is an overlap between the ground and excited state in the readout
voltage histogram. A threshold is chosen to digitize the readout
voltage $V$ to $+1$ for $|g\rangle$ and $-1$ for $|e\rangle$.
A readout error yields an incorrect interpretation of the qubit state ($g$
as $e$ or $e$ as $g$), and produces two consecutive errors in the inferred photon number as shown in Fig.\ \ref{fig:Fig2-example}(c).
When a $\pi$ pulse error
happens, the qubit is falsely flipped (unflipped) even though the
cavity is in the zero-photon (one-photon) state. Hence, $\pi$ pulse errors
have the same effect as qubit $T_{1}$ jumps. From now on, we do not
analyze $\pi$ pulse errors explicitly as they can be included in the $T_{1}$
error rate. From Fig.\  \ref{fig:Fig2-example}(c), it is clear that the experimental
imperfections can greatly hinder our ability to distinguish zero-photon
and one-photon states if we rely solely on correlating neighboring qubit
readouts.

In the following section, we develop a Bayesian smoothing algorithm
taking into account the entire measurement record to best estimate the
photon state at each interrogation. Fig.\  \ref{fig:Fig2-example}(d) shows the conditional probability of the one-photon
state obtained from applying the smoothing algorithm to the
example data in Fig.\  \ref{fig:Fig2-example}(b). Compared to Fig.\  \ref{fig:Fig2-example}(c), it
is clear that the errors introduced by $T_{1}$ jumps and readout
errors are very effectively corrected after filtering based on the entire measurement
record.

\subsection{Hidden Markov Model}
We are interested in the time history of photon number occupation in the cavity.   What we actually have access to is the time history of the qubit readout.  This measurement record is imperfectly related to the time history of the qubit state which in turn is imperfectly related to the photon number occupation history.  We are thus dealing with a three-layer \textit{hidden Markov model} \cite{Kobayashi12}.  Our goal is to develop a filter which optimally estimates the photon number occupation record in terms of the information available in the qubit readout record.

\begin{figure}[htb]
\centering
\includegraphics[width=0.48\textwidth]{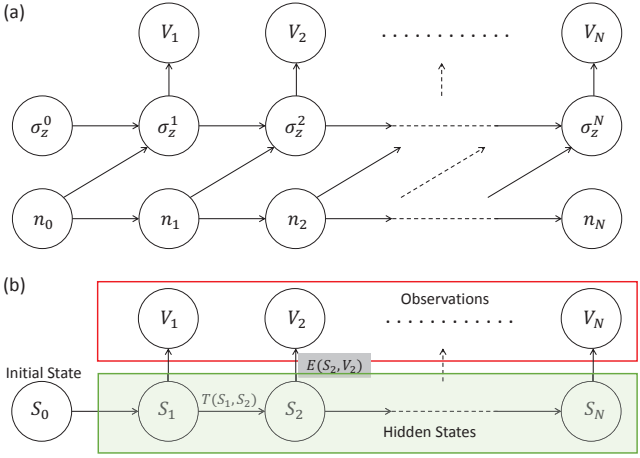}
\caption{(Color online) Hidden Markov model of the single-photon detector. (a) The Markov
chain for photon state $n$, qubit state $\sigma_{z}$ and readout
voltage $V$. (b) The simplified Markov chain after defining the joint
photon-qubit state $S=n\otimes\sigma_{z}$. Here, $S$ is the hidden
state and $V$ is the observation we collect in the experiment. The
transition between $S$ states is determined by the transition matrix
$\mathbf{\mathrm{T}}$ and the transition between $S$ and $V$ is governed by the emission
matrix $\mathbf{\mathrm{E}}$ [see Eq.\,(\ref{eq:T-E})].
}
\label{fig:Fig3-HMM}
\end{figure}

Because we are interested in the case that the mean cavity photon occupation
$\bar{n}$ is far below one, we neglect the high photon number states
and consider the zero-photon and one-photon cavity states only. In this
case, the evolution of the qubit-cavity system
can be represented by a Markov chain as illustrated in Fig.\,\ref{fig:Fig3-HMM}(a). The
cavity state evolution is a Markov chain dictated by photon jump probabilities $p_{01}$ from $0$ to $1$ and $p_{10}$ from $1$ to $0$ during each measurement duration $\tau_m$: $p_{10}\approx \kappa \tau_m$ and $p_{01}\approx \bar{n}\kappa \tau_m$, where $\kappa$ is the cavity line width and we assume that the time $\tau_m$ required to measure the qubit state is short compared to all other characteristic times of the system dynamics.
The qubit state is determined by the cavity photon number, the
qubit state from the previous interrogation, and qubit $T_1$ jump probabilities during $\tau_m$.
Without a $T_1$ jump, the qubit state is flipped (unflipped) if there is one-photon (zero-photon) in the cavity.  The qubit ends up in the opposite state in the presence of a $T_1$ jump.
 {
We distinguish the $T_1$ up and down events and the corresponding jump probabilities
are given by $p_{ge}\approx \Gamma_{\uparrow}\tau_m$ and $p_{eg}\approx \Gamma_{\downarrow}\tau_m$.
}%
Finally, the readout outputs the wrong qubit state with error probability $\varepsilon$.  (We assume the readout error probability is symmetric, i.e.\ is the same for both qubit states.)

By combining
the qubit state $\sigma_{z}^{i}$ and cavity state $n_{i}$ at each measurement $i$, we can
form a joint cavity-qubit state $S_{i}=n_{i}\sigma_{z}^{i}\in{0g,0e,1g,1e}$
and describe the process using a simplified Markov chain as shown
in Fig.\,\ref{fig:Fig3-HMM}(b). The transition between $S_i$ and $S_{i+1}$ is determined by the transition matrix element $\mathbf{\mathrm{T}}(S_i,S_{i+1})=P(S_{i+1}|S_i)$, which is the
conditional probability of $S_{i+1}$ given $S_i$.
The transition between $S_i$ and $V_i$ is determined by the emission matrix element $\mathbf{\mathrm{E}}(S_i,V_i)=P(V_i|S_i)$, which is the conditional probability of $V_i$ given $S_i$.
The transition matrix is determined
by the combination of photon jump probabilites $p_{01}$ and $p_{10}$, and qubit $T_1$ jump probabilities $p_{ge}$ and $p_{eg}$. The emission matrix depends solely on the readout error probability $\varepsilon$.
In particular, in the basis of $(0g,0e,1g,1e)$, the transition and emission matrices read
\begin{eqnarray}
    \mathbf{\mathrm{T}}&=&\begin{pmatrix}
       \tilde{p}_{01}\tilde{p}_{ge} & \tilde{p}_{01}p_{eg} & p_{10}p_{eg} & p_{10}\tilde{p}_{ge}\\
       \tilde{p}_{01}p_{ge} & \tilde{p}_{01}\tilde{p}_{eg} & p_{10}\tilde{p}_{eg} & p_{10}p_{ge}\\
       p_{01}\tilde{p}_{ge} & p_{01}p_{eg} & \tilde{p}_{10}p_{eg} & \tilde{p}_{10}\tilde{p}_{ge}\\
       p_{01}p_{ge} & p_{01}\tilde{p}_{eg} & \tilde{p}_{10}\tilde{p}_{eg} & \tilde{p}_{10}p_{ge}
      \end{pmatrix}, \nonumber \\
       \mathbf{\mathrm{E}} &=& \begin{pmatrix}
     \tilde{\varepsilon} & \varepsilon
  & \tilde{\varepsilon} &  \varepsilon  \\
    \varepsilon &   \tilde{\varepsilon}
   &  \varepsilon & \tilde{\varepsilon}
    \end{pmatrix},
\label{eq:T-E}
\end{eqnarray}
where $\tilde{p}=1-p$ and $\tilde{\varepsilon}=1-\varepsilon$. We will address the question of how to estimate those parameters in Sec. \ref{sec:PE-Robust}.

This defines the parameters of the hidden
Markov Model \cite{Kobayashi12}. The measurement records $V_{1}\cdots V_{N}$
are the $N$ \textit{observations} we have access to, and the qubit-cavity states
$S_{1}\cdots S_{N}$ are the \textit{hidden states} not directly seen by the experimenter.
We want to estimate the conditional probability of hidden state $S_{i}$
at each interrogation $i$ given the entire observed data $V_{1}\cdots V_{N}$:
$P(S_{i}|V_{1}\cdots V_{N})$, which yields $P(n_{i}|V_{1}\cdots V_{N})$
after summing over the qubit state $\sigma_{z}^{i}$. Given the transition
and emission matrices and the observed data, the conditional probability
of hidden states can be computed efficiently using the \textit{forward-backward
algorithm} \cite{Kobayashi12},
\begin{equation}
P(S_{i}|V_{1}\cdots V_{N})=\frac{P(S_{i}|V_{1}\cdots V_{i})P(V_{i+1}\cdots V_{N}|S_{i})}{\sum_{S_i}P(S_{i}|V_{1}\cdots V_{i})P(V_{i+1}\cdots V_{N}|S_{i})},
\label{eq:CondP}
\end{equation}
where $P(S_{i}|V_{1}\cdots V_{i})$ and $P(V_{i+1}\cdots V_{N}|S_{i})$ are the forward and backward probabilities, and can be calculated iteratively using the forward and backward algorithms
\begin{widetext}
\begin{subequations}
\begin{eqnarray}
  && P(S_{i}|V_{1}\cdots V_{i})=\frac{\sum_{S_{i-1}}E(S_i,V_i)T(S_{i-1},S_i)P(S_{i-1}|V_{1}\cdots V_{i-1})}{\sum_{S_i,S_{i-1}}E(S_i,V_i)T(S_{i-1},S_i)P(S_{i-1}|V_{1}\cdots V_{i-1})},\\
 && P(V_{i+1}\cdots V_{N}|S_{i})=\sum_{S_{i+1}} P(V_{i+2}\cdots V_{N}|S_{i+1}) E(S_{i+1},V_{i+1})T(S_i,S_{i+1}).
 \end{eqnarray}
 \label{eq:FB}
 \end{subequations}

\end{widetext}

Fig.\ \ref{fig:Fig2-example}(d) shows the conditional probability of the one-photon
state $P(n_{i}=1|V_{1}\cdots V_{N})$ obtained from applying the forward-backward algorithm to the
example data in Fig.\  2(b).
The result is very close to the true photon state sequence shown in \ref{fig:Fig2-example}(a).
We believe the smoothing algorithm is optimal in the sense that we make use of all the information
gathered from all observations to infer the photon state at each measurement.
Similar quantum state smoothing has been carried out before in several quantum systems \cite{GuerlinNat07,HumePRL07,GammelmarkPRL13,GammelmarkPRA14,GuevaraArXiv15}.
Next, we carry out a detailed analysis
of the performance of our single-photon detector.

\subsection{Fidelity, Detection Efficiency and Dark Count}
To quantify the performance of the single-photon detector, we generate
a random sequence of $5000$ photon jumps between the zero-photon and one-photon states.  These are drawn from the probability distribution associated with a cavity life time $\tau_c=1/\kappa=100T_\mathrm{m}$ and photon occupation $\bar{n}=0.01$.
%
%
 {
We choose $\tau_\mathrm{m}=300\,\mathrm{ns}$, $T_1=20\tau_\mathrm{m}=6\,\mathrm{\mu s}$, $\Gamma_\uparrow=0.1/T_1$, and $\varepsilon=0.01$ in accordance with recent circuit-QED experiments \cite{SunNat14}.
These parameters correspond to the assumption that (in equilibrium) there is a probability of $p=10\%$  to find the qubit in the excited state.
Based on the transition and emission probabilities, we generate a corresponding random instance of
the measurement record $V_{1}\cdots V_{N}$. Applying the smoothing algorithm to the measurement record
yields the one-photon conditional probability $P(n_i=1|V_{1}\cdots V_{N})$.
}

\begin{figure}[tbh]
\centering
\includegraphics[width=0.4\textwidth]{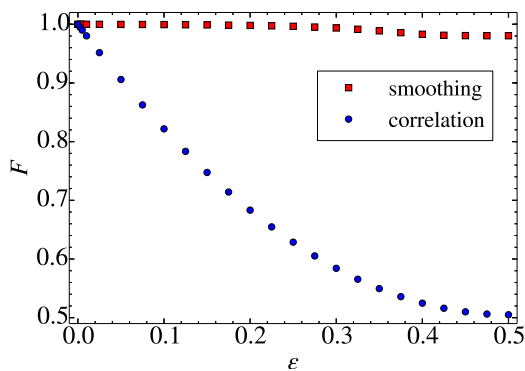}
\caption{(Color online)  Detector fidelity ($F$) as a function of readout error probability $\varepsilon$ obtained using the smoothing algorithm (red solid) and direct correlation (blue circle).
Here, we set $\tau_c=100\tau_m$, $T_1=20\tau_m$ and $\bar{n}=0.01$.
}
\label{fig:Fig35-Fidelity}
\end{figure}

First, we define the detector fidelity as the probability for correctly predicting the cavity photon state averaged over the whole measurement record
\begin{equation}
 F=\frac{\sum_{j=0,1}\sum_{i\in \mathbb{N}_{j}} P(n_i=j|V_{1}\cdots V_{N}) }{\sum_{j=0,1}\sum_{i\in \mathbb{N}_{j}}1},
\end{equation}
where $\mathbb{N}_{1}$ and $\mathbb{N}_{0}$ are the sets of measurements corresponding to one-photon and zero-photon states of the cavity, respectively.
From Fig.\,\ref{fig:Fig35-Fidelity}, we see that the smoothing algorithm is superior to the simple direct correlation method in faithfully predicting the cavity photon states.
 {We emphasize that at the special point $\varepsilon=0.5$, the qubit measurements are completely random and do not provide any useful information to infer the qubit state and the cavity photon number.
The fidelity of direct correlation goes to $50\%$ as expected for random guessing.
However, for the smoothing algorithm the fidelity is still close to $1$ at $\varepsilon=0.5$.
This is because we input the cavity mean photon occupation $\bar{n}=0.01$ into the smoothing algorithm which assigns one-photon or zero photon state to each observation with probability $1-\bar{n}$ or $\bar{n}$ (neglecting two or more photon states), leading to the near unity fidelity $F\approx (1-\bar{n})^2+\bar{n}^2=1-2\bar{n}+O(\bar{n}^2)$.
}


In order to better quantify the detector, we define separate figure of merits namely \textit{detection efficiency} $\eta$ (inefficiency $P_{10}$) for the one photon events and \textit{dark count probability} $P_{01}$ for the zero photon events
\begin{subequations}
\label{eq:eta-P01}
 \begin{eqnarray}
\eta &=&1-P_{10}=\frac{ \sum_{i\in \mathbb{N}_{1}}P(n_i=1|V_{1}\cdots V_{N}) } {\sum_{i\in \mathbb{N}_{1}}1}, \\
P_{01} &=&\frac{ \sum_{i\in \mathbb{N}_{0}}P(n_i=1|V_{1}\cdots V_{N}) } {\sum_{i\in \mathbb{N}_{0}}1},
\end{eqnarray}
\end{subequations}
$\eta$ ($P_{01}$) is the probability to detect a photon when there is one (zero) photon in the cavity.   {Because the filter correctly recognizes that photons live in the cavity for a mean residence time $\tau_c=1/\kappa$,} the dark count rate is simply given by $\gamma_D=P_{01}\kappa$. For comparison, we also
 present 
 $\eta$ and $P_{01}$ of the inferred photon states from direct correlations of neighboring measurements, as shown in Fig.\  \ref{fig:Fig2-example}(c).

\begin{figure}[tbh]
\centering
\includegraphics[width=0.5\textwidth]{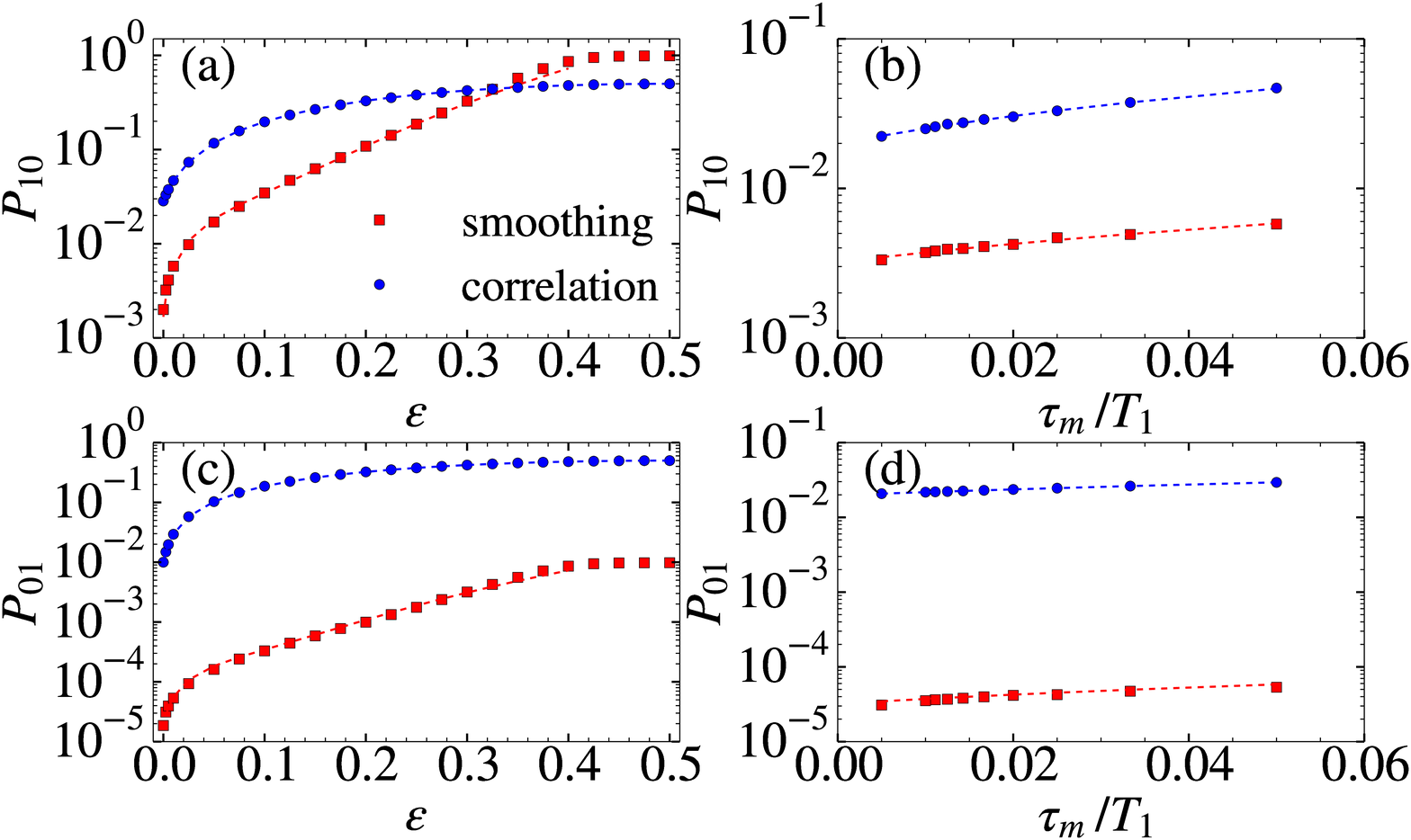}
\caption{(Color online) Detection inefficiency $P_{10}$ and dark count probability $P_{01}$. (a)\&(c) $P_{10}$ and $P_{01}$ as a function of
readout error probability $\varepsilon$ obtained using the smoothing algorithm (red solid) and direct correlation (blue circle). Here, $T_1=20\tau_m$.
(b)\&(d) $P_{10}$ and $P_{01}$ as a function of $\tau_m/T_1$. Here, we set $\varepsilon=0.01$.
In all the plots, we have $\tau_c=100\tau_m$ and $\bar{n}=0.01$.
The dashed lines are fits to models. See main text for details.
}
\label{fig:Fig4-P10-P01}
\end{figure}

Fig.\ \ref{fig:Fig4-P10-P01} shows $P_{10}$ and $P_{01}$ as a function of $\varepsilon$ and $\tau_m/T_1$.
Three comments are in order.
First, it is clear that except for very large $\varepsilon$, the smoothing algorithm gives an approximately \textit{one order-of-magnitude} lower detector inefficiency and \textit{three orders-of-magnitude} lower dark count than using direct correlations.
This confirms that it is superior to process the entire measurement data altogether than to use only two neighboring measurements to infer photon states.
Our smoothing algorithm makes use of \textit{nonlocal correlations} among photon states: on average the cavity state stays unchanged within the cavity life-time.  {The method produces the most improvement in the case where} it is \textit{much more likely} to have $T_1$ and readout errors than a short-lived photon, namely
\begin{equation}
p_{01}p_{10}=\bar{n}(\kappa \tau_m)^2\ll \text{min}(\tau_m/T_1, \varepsilon),
\label{eq:Key-Cond}
\end{equation}
where $p_{01}$ and $p_{10}$ are the probabilities for a photon jump into and out of the cavity during a single measurement $\tau_m$, respectively.
We stress that this condition underlines the \textit{key working principle} behind the smoothing algorithm.
As long as this criterion is met, the details of the system are largely irrelevant.
For current circuit-QED systems, this condition is very well satisfied \cite{SunNat14}.
For instance, with $\tau_\mathrm{c}=100\tau_\mathrm{m}=30\,\mathrm{\mu s}$, $\bar{n}=0.01$, $T_1=20\tau_\mathrm{m}=6\, \mathrm{\mu s}$, $\varepsilon=0.01$, $\tau_\mathrm{m}/T_1=0.05$ while $p_{01}p_{10}=\bar{n}(\tau_\mathrm{m}/\tau_\mathrm{c})^2=10^{-6}$.
With this built-in \textit{prior knowledge}, the smoothing algorithm is able to reject \textit{artificial} short-lived photon events created by individual $T_1$ jumps or readout errors and to recover the photon state sequence to a high accuracy.
For very large $\varepsilon$, concatenated errors have an increased chance to occur, creating artificial long-lived photon events.
The smoothing algorithm fails to identify those errors and the detector performance degrades for large $\varepsilon$ as shown in Fig.\,\ref{fig:Fig4-P10-P01}(a)-(b).
In particular, the detector efficiency from the smoothing algorithm becomes worse than that of the direct correlation method.
This is due to the asymmetry between photon jump in and jump out rates.
In this case, the cavity mean photon number is much smaller than $1$, and hence the smoothing algorithm favors the $0$-photon events and underestimates the cavity occupation.
However, we stress that the overall fidelity from the smoothing algorithm is still much better than the fidelity from the direct correlation as shown in Fig.\,\ref{fig:Fig35-Fidelity}.   {This is because the filter is designed to optimize fidelity, not efficiency.}

Second, by comparing Fig.\,\ref{fig:Fig4-P10-P01}(a) and (c) to Fig.\,\ref{fig:Fig4-P10-P01}(b) and (d),
we find that both $P_{10}$ and $P_{01}$ have a \textit{stronger dependence} on $\varepsilon$ than $T_1$.
This is because each $T_1$ jump creates an artificial photon event lasting for only \textit{one} measurement step,
while a readout error generates an artificial photon event lasting for \textit{two} measurement steps.
Therefore, $T_1$ jumps are relatively easier to correct than the readout errors, and hence are less of a limiting factor.
This implies that to improve the detector performance, experimental efforts should be directed toward improving readout fidelity rather than suppressing $T_1$ jumps.

Finally, the dashed lines in Fig.\,\ref{fig:Fig4-P10-P01} show model fits (described below) to the numerical results with a very good agreement.
For the direct correlation method, by combining the $T_1$ jump probabilities with the readout errors, we obtain
\begin{subequations}
\label{eq:fit-corr}
\begin{eqnarray}
 P^C_{10}&=&[\varepsilon^2+(1-\varepsilon)^2]\frac{p_{eg}+p_{ge}}{2}\nonumber\\
 &+&2\varepsilon(1-\varepsilon)(1-\frac{p_{eg}+p_{ge}}{2}), \\
 P^C_{01}&=&[\varepsilon^2+(1-\varepsilon)^2]\frac{2p_{eg}p_{ge}}{p_{eg}+p_{ge}}\nonumber\\
 &+&2\varepsilon(1-\varepsilon)(1-\frac{2p_{eg}p_{ge}}{p_{eg}+p_{ge}}).
 \end{eqnarray}
 \end{subequations}
Eqs.\,(\ref{eq:fit-corr}a,b) can be understood by noticing that for any pair of neighboring measurements, the correlation between them gives a false photon count when: 1) both readouts are right or wrong [with probability $\varepsilon^2+(1-\varepsilon)^2$], and a qubit $T_1$ jump happens in one of the two measurements;
2) one of the readouts is wrong [with probability $2\varepsilon(1-\varepsilon)$] and the qubit has either no $T_1$ jumps or experiences $T_1$ jumps in both measurements.
The above two cases correspond to the first and second terms in Eq.\,(\ref{eq:fit-corr}a,b), respectively.

For the smoothing algorithm, we fit the numerical result to the following model which is valid for samll $\varepsilon$ and $\tau_m/T_1$
\begin{subequations}
\label{eq:fit-smooth}
\begin{eqnarray}
  && P_{10}= \frac{\tau_m}{\tau_c} f\big(\varepsilon,\frac{\tau_m}{T_1}\big), \\
   && P_{01}= \frac{\tau_m}{\tau_c/\bar{n}} f\big(\varepsilon,\frac{\tau_m}{T_1}\big),
\end{eqnarray}
\end{subequations}
where $f(\varepsilon,\tau_m/T_1)$ is a polynomial function of $\varepsilon$ and $\tau_m/T_1$ further described below.
The same $f(\varepsilon,\tau_m/T_1)$ function is used to fit $P_{01}$ according to Eq.\,(\ref{eq:fit-smooth}b).

The fact that the same polynomial function describes both $P_{10}$ and $P_{01}$ indicates that they have the same physical origin.
Let us consider the one-photon events.
When an isolated $T_1$ or readout error occurs in the middle of the one-photon sequence,
the smoothing algorithm is capable of rejecting this error because it takes into account both the measurements before and after it, and the majority of the measurements indicates that the cavity has one photon.
In contrast, when an error happens at the very \textit{edge} of a one-photon sequence, i.e., right after a photon jumps into the cavity,
this error can be interpreted as either the \textit{tail of the zero-photon sequence} or an actual \textit{photon jump} accompanied by a $T_1$ or readout error.
Since the former interpretation is much more likely, the smoothing algorithm will falsely identify the cavity state as a zero-photon state with a probability close to $1$.
Therefore, errors at the edges are the \textit{dominant source} of detection inefficiency $P_{10}$.
The probability for an error to occur at each measurement is determined by $\varepsilon$ and $\tau_m/T_1$, and we have to consider
all possible scenarios of how concatenated errors could occur at the edges.
After summing over all the contributions from the edges, we expect to obtain a polynomial function $f(\varepsilon,\tau_m/T_1)$.
And the same argument applies to the dark count probability $P_{01}$ and hence we expect to have the same polynomial function $f(\varepsilon,\tau_m/T_1)$.

To obtain the detection inefficiency and dark count probability, we divide $f(\varepsilon,\tau_m/T_1)$ by the average number of measurements of one-photon and zero-photon events, which are simply
$\tau_c/\tau_m$ and $\tau_c/(\bar{n}\tau_m)$. This leads to Eq.\,(\ref{eq:fit-smooth}).
The analytical form of $f(\varepsilon,\tau_m/T_1)$ is difficult to obtain because it requires summing over all possible
scenarios of errors with each scenario weighted by the probability for the smoothing algorithm to falsely interpret the photon state.
 { Instead, in the following we do separate fits of $P_{10}$ and $P_{01}$ against the variables $\varepsilon$, $\tau_m/T_1$, $\tau_m/\tau_c$ and $\bar{n}$ to verify the formula in Eq.\,(\ref{eq:fit-smooth}). }

\begin{figure}[tbh]
\centering
\includegraphics[width=0.45\textwidth]{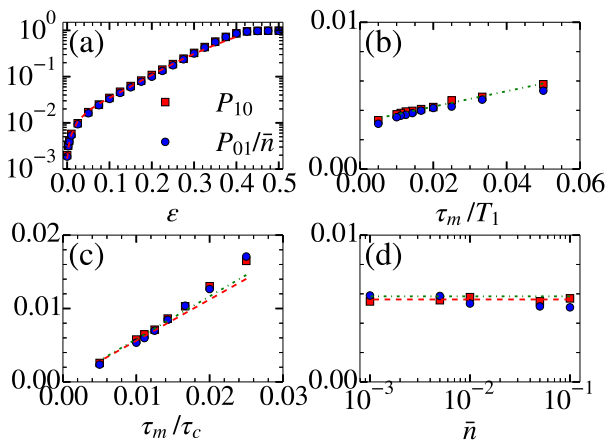}
\caption{(Color online) Fitting to the model.
$P_{10}$ (red square) and $P_{01}/\bar{n}$ (blue circle) as a function of (a) $\varepsilon$, (b) $\tau_m/T_1$, (c) $\tau_m/\tau_c$ and (d) $\bar{n}$.
The lines in (a) and (b) are fits to the model in Eq.\,(\ref{eq:fit-smooth})
The red dashed and green dotted lines in (c) and (d) are self-consistent fits with coefficients taken from the fits in (a) and (b), respectively.
The parameters are the same as in Fig.\,\ref{fig:Fig4-P10-P01}}
\label{fig:Fig5-Scaling}
\end{figure}

\begin{figure*}[tb]
\centering
\includegraphics[width=0.85\textwidth]{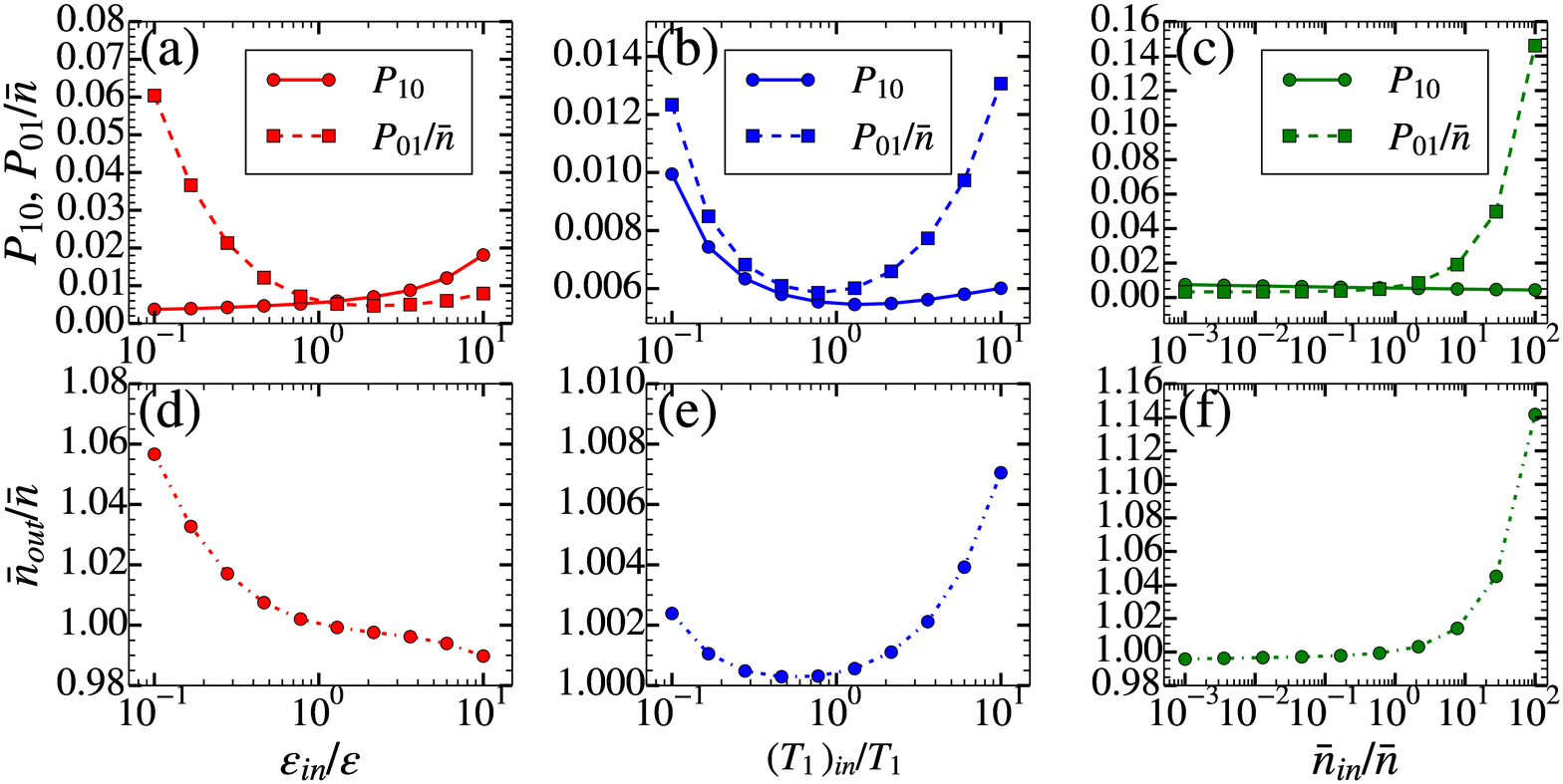}
\caption{(Color online) Robustness against input parameters. (a)-(c) $P_{10}$ (circle) and $P_{01}/\bar{n}$ (square) as a function of $\varepsilon_{\text{in}}/\varepsilon$, $(T_1)_{\text{in}}/T_1$ and $\bar{n}_{\text{in}}/\bar{n}$.
(d)-(f) $\bar{n}_{\text{out}}/\bar{n}$ as a function of $\varepsilon_{\text{in}}/\varepsilon$, $(T_1)_{\text{in}}/T_1$ and $\bar{n}_{\text{in}}/\bar{n}$.
The lines are a guide to eye.
For each plot, the input parameters are the same as their actual values except the varying parameter. The actual parameters values are set to be $T_1=20\tau_m$, $\varepsilon=0.01$, and $\bar{n}=0.001$.
}
\label{fig:Fig6-Robustness}
\end{figure*}

Eq.\,(\ref{eq:fit-smooth}) features four unique dependences on the system parameters:

1) The detection inefficiency and the dark count probability are connected via the simple relation
\begin{equation}
 P_{10}= P_{01}/\bar{n}.
 \label{eq:P10-P01}
\end{equation}
\textit{This is a unique property of our detector, not characteristic of generic single-photon detectors.}
Eq.\,(\ref{eq:P10-P01}) comes about because both the quantum inefficiency $P_{10}$ and the dark count $P_{01}$
have dominant contributions from the errors at the falling and rising edges between zero- and one-photon states, while the zero-photon states have an average life time $1/\bar{n}$ times of that for the one-photon states.

In Fig.\,\ref{fig:Fig5-Scaling}(a)-(d), we plot $P_{10}$ together with $P_{01}/\bar{n}$ as a function of $\varepsilon$, $\tau_m/T_1$ and $\tau_m/\tau_c$.
Indeed, $P_{10}$ is identical to $P_{01}/\bar{n}$ within a few percent.

2) Both $P_{10}$ and $P_{01}/\bar{n}$ are polynomial functions of $\varepsilon$ and $\tau_m/T_1$.
In Fig.\,\ref{fig:Fig5-Scaling}(a), we fit $f(\varepsilon,\tau_m/T_1)$ to the third order function $h(\varepsilon)\equiv f(\varepsilon,\tau_m/T_1=0.05)=0.17+41.16\varepsilon-227.43\varepsilon^2+1444.93\varepsilon^3$.
In Fig.\,\ref{fig:Fig5-Scaling}(b), a first-order fit of $g(\tau_m/T_1)\equiv f(\varepsilon=0.01,\tau_m/T_1)=0.32+5.25(\tau_m/T_1)$ gives a good agreement with the numerical simulation.

3) Both $P_{10}$ and $P_{01}/\bar{n}$ depend on $\tau_m/\tau_c$ linearly. Again, this is a unique feature of our detector.
To be self-consistent, we use the polynomial functions $h(\varepsilon)$ and $g(\tau_m/T_1)$ from the fits to $\varepsilon$ and $\tau_m/T_1$,
and set $\varepsilon=0.01$ and $T_1=20\tau_m$ to obtain the coefficients $c_1=h(\varepsilon=0.01)$ and $c_2=g(\tau_m/T_1=0.05)$ of the linear dependence on $\tau_m/\tau_c$.
Fig.\,\ref{fig:Fig5-Scaling}(c) shows that those parameter-free fits $c_1(\tau_m/\tau_c)$ and $c_2(\tau_m/\tau_c)$ agree reasonably well with the numerical results.

4) $P_{10}$ and $P_{01}/\bar{n}$ are independent of $\bar{n}$.
Similar to 3), a self-consistent fit can be done by taking the fitting functions $h(\varepsilon)$ and $g(\tau_m/T_1)$ in 2) and setting $\varepsilon=0.01$, $T_1=20\tau_m$ and $\tau_c=100\tau_m$.
As shown in Fig.\,\ref{fig:Fig5-Scaling}(d), the values obtained from the fits are in good agreement with the numerical results.


\subsection{Parameter Estimation and Robustness of the Detector Performance}
\label{sec:PE-Robust}

Our smoothing algorithm takes the transition and emission matrices in Eq.\,(\ref{eq:T-E}) as inputs to process the gathered data.
These two matrices depend on five system parameters: readout error $\varepsilon$, the qubit transition rates $\Gamma_{\uparrow,\downarrow}$, 
the cavity life time $\tau_c=1/\kappa$ and the mean photon occupation $\bar{n}$.
In previous plots, we simply assumed we knew those parameters and used their actual values in the smoothing algorithm.
In practice, however, we have to ask ourselves: \textit{how do we estimate those parameters?}
For this question, an initial estimation of the system parameters is sufficient and can be done using standard benchmark experiments in circuit-QED \cite{PaikPRL11, SunNat14}.
For instance, the qubit transition rates can be obtained from the mean excited state population, time domain measurements of the natural dynamics and/or by applying a $\pi$ pulse and monitoring the transient population decay back to equilibrium \cite{PaikPRL11,SunNat14}.
 To calibrate the readout error $\varepsilon$, we can first prepare the qubit in either $|g\rangle$ or $|e\rangle$ and then measure the corresponding histogram of readout voltage (effectively computing the repeatability of the state measurement) \cite{SunNat14}.
$\tau_c$ can be estimated by preparing the cavity in $n=1$ photon state and monitoring its population decay \cite{SunNat14}.
Lastly, the cavity mean photon occupation is very nearly equal to the probability of $n=1$ state when $\bar n \ll 1$.
A measurement of the photon number distribution will give us the required information \cite{JohnsonNatPhys10,SunNat14}.
The parameters extracted through the benchmark measurements will be our input parameters to the smoothing algorithm. We denote them as $\bar{n}_{\text{in}}$, $\varepsilon_{\text{in}}$ etc.
We reserve $\bar{n}$, $\varepsilon$ etc. for the actual values of those parameters.

Various experimental imperfections will inevitably lead to \textit{deviations} of the estimated input parameters from their actual values.
More importantly, during the operation of the single-photon detector, all of these parameters might experience small \textit{drifts}.
This naturally leads us to ask the question: \textit{how robust is our detector against variations of the input parameters?}

\begin{figure}[htb]
\centering
\includegraphics[width=0.5\textwidth]{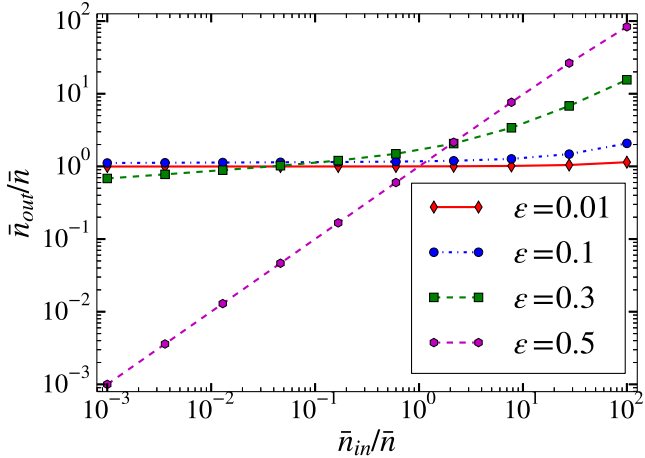}
\caption{(Color online) The extracted output mean photon number $\bar{n}_{\text{out}}$ as a function the input mean photon number $\bar{n}_{\text{in}}$ for $\varepsilon=0.01,0.1,0.3,0.5$.
 Here, $\tau_c=100\tau_m$, $T_1=20\tau_m$ and $\bar{n}=0.001$.
}
\label{fig:Fig_nout_nin}
\end{figure}

Fig.~\ref{fig:Fig6-Robustness} shows $P_{10}$ and $P_{01}/\bar{n}$ as a function of $\varepsilon_{\text{in}}/\varepsilon$, $(T_1)_{\text{in}}/T_1$ and $\bar{n}_{\text{in}}/\bar{n}$.
In each plot, except for the varying input parameter, all the other input parameters are set to be their actual values.
In general, we observe that the dark count $P_{01}/\bar{n}$ is more sensitive to the inaccuracy of input parameters than the detection inefficiency $P_{10}$.
This is because zero-photon sequences are about $1/\bar{n}$ times longer than one-photon sequences on average.
When the input parameters match their actual values, the errors occurring at the edges are the dominant source of both dark counts and detection inefficiency as shown in Eq.\,(\ref{eq:fit-smooth}).
However, as the parameter deviations increase, the smoothing algorithm becomes less capable of detecting errors.
The longer zero-photon sequences have more $T_1$ and $\varepsilon$ errors and hence are more vulnerable to the inaccuracy of input parameters than the one-photon sequences.
In particular, we notice that $P_{01}/\bar{n}$ can be as high as $\sim 15\%$ when $\bar{n}_{\text{in}}=100\bar{n}$ as shown in Fig.\,\ref{fig:Fig6-Robustness}(c).
 {
In Fig.~\ref{fig:Fig_nout_nin} we examine the variation of the extracted output mean photon number for cases with different readout error $\varepsilon$.
The performance degrades as $\varepsilon$ increases as there is less useful information from the observations.
In the limit of $\varepsilon=0.5$, the observations are completely useless in inferring the cavity state.
The extracted output mean photon occupation $\bar{n}_{\text{out}}$ is simply the same as the input photon occupation $\bar{n}_{\text{in}}$ we pass to the smoothing algorithm.
In practice, it is unlikely that dark counts would cause a miscalibration of $\bar n$ by a factor of 100 or $\varepsilon$ on the order of $0.5$.
But for the detection of the extremely weak axion signal, we want to maintain the highest possible sensitivity of the detector while minimizing the dark count rate.
}

\begin{figure}[htb]
\centering
\includegraphics[width=0.45\textwidth]{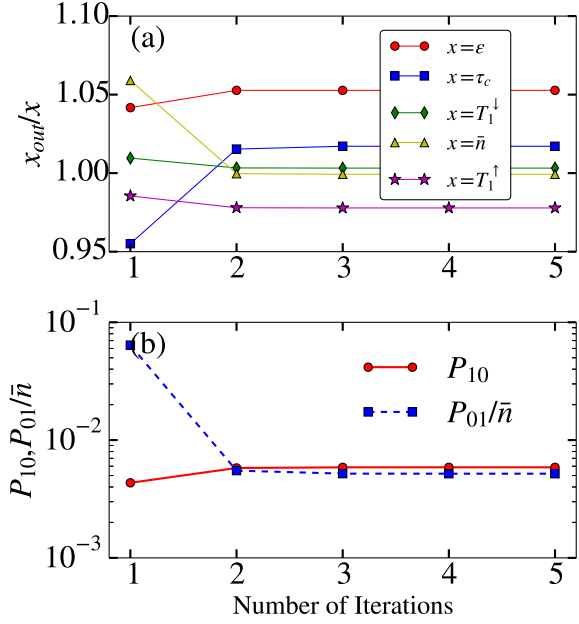}
\caption{(Color online) Convergence of the iterative procedure. (a) The ratio between extracted parameters $x_{\text{out}}$ and their actual values $x$ as a function of number of iterations. Here, we show the cases of $x=\varepsilon$, $\tau_c$, $T_1^{\downarrow}$, $\bar{n}$ and $T_1^{\uparrow}$ with input values $\varepsilon_{\text{in}}=0.1\varepsilon$,
 {$\Gamma_{\uparrow,\downarrow,\mathrm{in}}=0.1\Gamma_{\uparrow,\downarrow}$, $(\tau_c)_{\mathrm{in}}=10\tau_c$ and $\bar{n}_{\mathrm{in}}=100\bar{n}$.}
%
 (b) $P_{10}$ and $P_{01}/\bar{n}$ as a function of number of iterations.
 Here, $\varepsilon=0.01$, $\tau_c=100\tau_m$, $T_1=20\tau_m$ and $\bar{n}=0.001$.
}
\label{fig:Fig7-Conv}
\end{figure}

To systematically eliminate the impact of inaccurate parameter estimations and potential parameter variations, we develop an \textit{iterative procedure} to \textit{extract} the system parameters and improve the detector performance.
The first step is to run the smoothing algorithm with the input parameters estimated from the benchmark experiments.
Next, we can \textit{reconstruct} the photon and qubit states from the output data.
This can be done by applying a 
 {threshold} to the conditional probability in Eq.\,(\ref{eq:CondP})  {to separate one-photon events from zero-photon events}.
With the reconstructed states, we can easily extract all of the system parameters and use them as input to run the smoothing algorithm iteratively.

To test the \textit{robustness} of the iterative procedure,
we choose input parameters to be very different from their actual values: $\varepsilon_{\text{in}}=0.1\varepsilon$, 
 $\Gamma_{\uparrow,\downarrow,\mathrm{in}}=0.1\Gamma_{\uparrow,\downarrow}$, $(\tau_c)_{\mathrm{in}}=10\tau_c$ and $\bar{n}_{\mathrm{in}}=100\bar{n}$.
According to Fig.~\ref{fig:Fig6-Robustness}, this set of input parameters is expected to give us detection inefficiency and dark count on the level of $10\%$.
We define the extracted parameters after each iteration as $\varepsilon_{\text{out}}$ and so on, and compare them to their true values $\varepsilon$ etc.
Fig.~\ref{fig:Fig7-Conv}(a) shows that it takes only \textit{two iterations} for the extracted parameters to \textit{converge} to their actual values with an accuracy of a few percent.
Given that all of the initial input parameters are orders of magnitude off, such a rapid convergence is remarkable.
Correspondingly, Fig.\,\ref{fig:Fig7-Conv}(b) shows that after two iterations both $P_{10}$ and $P_{01}/\bar{n}$ quickly drop to $\sim5\times 10^{-3}$ reaching the values for perfectly accurate input parameters.
This demonstrates that with the iterative procedure our single-photon detector is \textit{immune} to inaccurate parameter estimations and potential parameter variations during the operation of axion searches.
This robustness is a positive feature of this filter for the detection of the extremely weak axion signal.


\section{Comparison between Single-Photon Detection and Linear Amplification}

In this section, we compare the single-photon detector to detection based upon linear amplification, and
show that the proposed single-photon detector can outperform linear amplifiers by several orders of magnitude in SNR. As described in Sec.~\ref{sec:QntSqzApp}, some of this deficit can be overcome with squeezing.

For detection using linear amplification, the SNR for an axion signal resonant with the cavity is determined from the Dicke radiometer equation \cite{Dicke1946}
\begin{equation}
 \alpha_{\text{SNR,la}}=\frac{ \eta\nac\kc }{{\namp+ \nt + \frac{1}{2}}}\sqrt{\frac{\tau}{B_w}},
 \label{eq:SNR-LA1}
\end{equation}
where $\eta$ is the efficiency with which a signal reaches the linear amplifier,  {$B_w$ is the axion signal bandwidth} and $\nac = 4 \ka B_w \na / \kappa^2$ is the cavity mean photon occupation due to an axion field resonant with the cavity. The factor $\eta\bar{n}_\mathrm{a}\kc$ is simply the signal power (in units of photons/s) that reaches the amplifier; whereas, the factor ${\namp + \nt + \frac{1}{2}}$ is the total noise power-spectral-density (in units of photons/[s$\cdot$Hz]) at the amplifier input (neglecting the tiny axion contribution).
For an axion search that operates at temperature $T \ll \hbar\wc/k_B$ and uses both a critically-coupled cavity $\kc = \kappa/2$ and a quantum-limited linear phase-preserving amplifier ($\namp = 1/2$),  {the thermal noise is negligible relative to the quantum noise and thus} Eq.\,(\ref{eq:SNR-LA1}) becomes
\begin{equation}
  \alpha_{\text{SNR,la}}= \frac{\eta\bar{n}_\mathrm{a}\kappa}{2} \sqrt{\frac{\tau}{B_w}}.
 \label{eq:SNR-LA2}
\end{equation}


In order to find the SNR for the single-photon detector, we note that in an observation time $\tau$, $N$ thermal photons are detected when axions are not present
\begin{equation}
 N=[\nt (1-P_{10})+(1-\nt)P_{01}]\kappa\tau,
\end{equation}
where the first and second terms correspond to the detection of thermal photons with occupation $\nt$ and the dark counts, respectively.  {The final factor $\kappa\tau$ represents the number of statistically independent measurements. (Note that it is $\kappa$ not $\kc$ which appears, because the unlike the case of the linear amplifier, the photons are detected \emph{inside} the cavity.)}
When the cavity is tuned into resonance with axions, there are excess photons due to conversion from axions
\begin{equation}
 \tilde{N}=[(\nt+\bar{n}_\mathrm{a})(1-P_{10})+(1-\nt-\bar{n}_\mathrm{a})P_{01}]\kappa \tau.
\end{equation}
Assuming that $N$ and $\tilde{N}$ are large enough to apply Gaussian statistics, the SNR of the single photon detector is given by
\begin{equation}
  \alpha_{\mathrm{SNR,spd}}=\frac{\tilde{N}-N}{\sqrt{ \tilde{N}} }.
\end{equation}
Using Eq.\,(\ref{eq:P10-P01}): $P_{01}=P_{10}\bar{n}$ which is valid when $\varepsilon$ and $\tau_m/T_1$ are small, we can simplify the SNR as
\begin{equation}
 \alpha_{\mathrm{SNR,spd}}=\frac{\bar{n}_\mathrm{a} \sqrt{\kappa \tau}}{\sqrt{(\nt+\bar{n}_\mathrm{a})}},
\label{eq:SNR-SPD3}
 \end{equation}
where we have neglected the second order contribution in both $\nt$ and $\bar{n}_\mathrm{a}$ as we assume $\nt, \bar{n}_\mathrm{a}\ll 1$.
Within this approximation, our single photon detector effectively has a $100\%$ quantum efficiency and zero dark count. This ideal behavior is possible because our single photon counter makes repeated QND measurements of the photon number \textit{inside} the axion cavity. Now, the only limiting factor is the photon counting shot noise due to thermal photons and axion photons.
Although the SNR for linear amplifiers is independent of the physical temperature as long as $\nt \ll \namp$, the SNR for the single-photon detector can be dramatically increased by reducing the thermal occupation $\nt$. As expressed in Eq.~(\ref{eq:IntegrationTime}), the axion scan rate is proportional to the square of the SNR. Thus using Eqs.\,(\ref{eq:SNR-LA2}) and (\ref{eq:SNR-SPD3}) the ratio of scan rates for the two types of detectors is
\begin{equation}
 \left(\frac{\alpha_{\mathrm{SNR,spd}}}{\alpha_{\text{SNR,la}}}\right)^2=\frac{2}{\eta_{\text{2}}^2}\frac{Q_\mathrm{c}}{\pi Q_\mathrm{a}}\frac{1}{(\nt+\bar{n}_\text{a})},
\label{eq:SNR-ratio}
 \end{equation}
where $Q_\mathrm{a}= \wc/(2\pi B_w) \sim 10^6$ is the $Q$ of the axion signal and $Q_\mathrm{c}= \wc/\kappa$ is the $Q$ of the axion cavity. As pointed out in Ref.\,\cite{LamoreauxPRD13}, single-photon detectors and linear amplifiers differ in two key ways.
First, for linear amplifiers the relevant bandwidth is the axion linewidth $B_w$ while for single-photon detectors it is the entire cavity linewidth $\kappa$.
This accounts for the factor $2\pi Q_\mathrm{a}/Q_\mathrm{c}\sim (2\pi)(10 - 100)$ in Eq.\,(\ref{eq:SNR-ratio}). Second, the noise floor for a quantum-limited phase-preserving linear amplifier is one photon per band (i.e.\ per Hz-sec) while the noise floor for the single-photon detector is the shot noise of cavity photons.
This introduces the factor $(\nt + \bar{n}_{\text{a}})$ in Eq.\,(\ref{eq:SNR-ratio}), which can be significantly less than $1$ at low temperatures. Without squeezing, single-photon detectors out perform linear amplifiers if $(\nt + \bar{n}_{\text{a}})< 2Q_\mathrm{c}/(\pi \eta_2^2Q_\mathrm{a})$.

As described in section~\ref{sec:SerRtEnh}, squeezing does not improve the SNR for detection of an axion signal on resonance, but it does increase the scan rate. If it is possible to transport microwave fields losslessly within an axion experiment ($\eta_1=\eta_2=1$), squeezing improves the scan rate of linear amplification by a factor $2G_1$ (Fig.~\ref{fig:SqueezBenA}a).  The condition for single photon detectors to outperform linear amplifiers is then more stringent, requiring $(\nt + \bar{n}_{\text{a}})< Q_\mathrm{c}/(G_1\pi \eta_2^2Q_\mathrm{a})$. In practice, efficiencies greater than 90\% seem technically remote, which limits the benefit of squeezing to less than a factor of 4 improvement in scan rate.

\begin{figure}[tbh]
\includegraphics[width=0.5\textwidth]{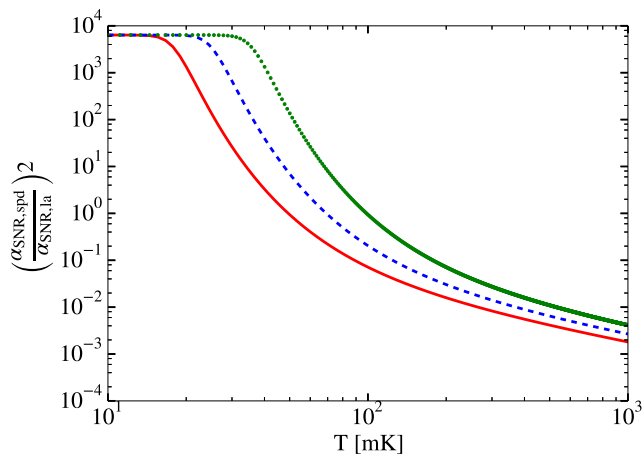}
\caption{(Color online) Comparison between the single-photon detector and linear amplifier. The ratio of scan rates is plotted as a function of temperature for three different axion cavity frequencies: 5~GHz (red solid), 7 GHz (blue dashed), 10 GHz (green dotted). Here, $\eta=1$, $Q_\mathrm{c}=Q_\mathrm{a}/100$, and $\bar n_\mathrm{a}=10^{-6}$. At the lowest temperatures, the single photon detector SNR saturates when $\nt<\bar{n}_\mathrm{a}$ and the detector is limited by the shot noise of the axion-derived photons.
}
\label{fig:SPDversusLA}
\end{figure}


Fig.~\ref{fig:SPDversusLA} shows the ratio of scan rates between our single-photon detector and an ideal quantum-limited linear amplifier as a function of temperature.
We plot this ratio for three different axion cavity frequencies ($\wc/(2\pi)=5$~GHz, 7~GHz, 10~GHz, and assuming $\bar{n}_\mathrm{a}=10^{-6}$, $Q_\mathrm{c}=Q_\mathrm{a}/100$. For any reasonable ratio $Q_\mathrm{a}/Q_\mathrm{c}$, photon counting is always the better strategy at sufficiently low temperature or high frequency because of the exponential suppression of thermal photons with $\hbar \wc/(k_B T)$ \cite{LamoreauxPRD13,ShokairIJMPA14}.
 {For the chosen parameters and at low temperatures on the order of $10$~mK, Fig.~\ref{fig:SPDversusLA} shows a speedup in scan rate of nearly $10^4$ using a single-photon detector compared to a linear amplifier without squeezing. The saturation of speedup is simply due to our decision to include the shot noise of axion-derived photons in the denominator of Eq.~(\ref{eq:SNR-ratio}).  This  becomes the dominant effect once the thermal and dark count rate is small enough.  Our Gaussian statistics analysis of the SNR remains valid in this limit provide the time interval is long enough that many axion-derived photons are observed.  However, in this ideal limit, even a single click of the detector during an appropriately chosen (shorter) time window indicates a high probability (obtained from Poisson rather than Gaussian statistics) of the presence of axions.
If however experimental constraints preclude operation at sufficiently low temperature to prefer single-photon detection, squeezing can speed up an axion search over what can be achieved with standard quantum-limited amplification.}

Finally, we note that if a plausible axion signal is identified by a single-photon detector, one should then switch to a linear amplifier.  Importantly, one can use the linear amplifier to verify that the axion signal has the expected bandwidth and the location of the spectral feature remains fixed as the amplifier  bandwidth is scanned across it.

\section{Conclusion}

The search for dark matter axions is a microwave measurement challenge in which intrinsic quantum fluctuations form the limiting background in the experiment. As such, newly developed quantum technologies that circumvent quantum noise are well poised to accelerate axion searches. In particular, we have analyzed in detail two concepts of quantum enhanced detection, namely quantum squeezing using Josephson parametric devices and QND single photon detection using transmon qubits. Squeezing can increase several fold the rate at which axion parameter space is searched, limited by the loss induced by transporting microwave fields within an experiment. Furthermore the squeezing concept is beneficial even when thermal noise dominates quantum noise and it is technically ready to use in existing axion searches. In contrast, transmon qubits have yet to be deployed in an axion search but quantum non-demolition photon detection can increase the axion search rate by several orders of magnitude. Indeed, for axion searches which operate at sufficiently high frequency, microwave photon detection will be necessary.

Here we have considered one particular application for which new microwave quantum technology is particularly well suited. In fact, photon detectors and squeezers have been crucial in the development of quantum optics and applications in quantum information processing.
For example, the detection of two coinciding photons heralds the entanglement between remote trapped ions \cite{MoehringNat07}. Similarly, two-mode squeezers create highly entangled optical clusters states \cite{Roslund2013}. We anticipate that quantum technologies for microwave measurement and control will find many more applications in quantum information science, precision measurement, and astrophysical instrumentation.

\begin{acknowledgments}
We would like to thank Ben Brubaker, Dan Palken, Uri Vool, Michael Hatridge, Anirudh Narla, Michel Devoret, Robert Schoelkopf, Mazyar Mirrahimi and Liang Jiang for fruitful discussions.
This work was supported by the National Science Foundation under Grants No.~1125844, PHY-1607223, and DMR-1301798, by the Army Research Office under ARO W911NF1410011 and by the Heising-Simons Foundation. A portion of this manuscript was written while KWL was hosted by the Yale Quantum Institute.
\end{acknowledgments}

\appendix

\section{Full two-mode squeezing with loss}\label{sec:App:FullTMSwithLoss}
\begin{figure*}[htb]
  \centering
  \includegraphics{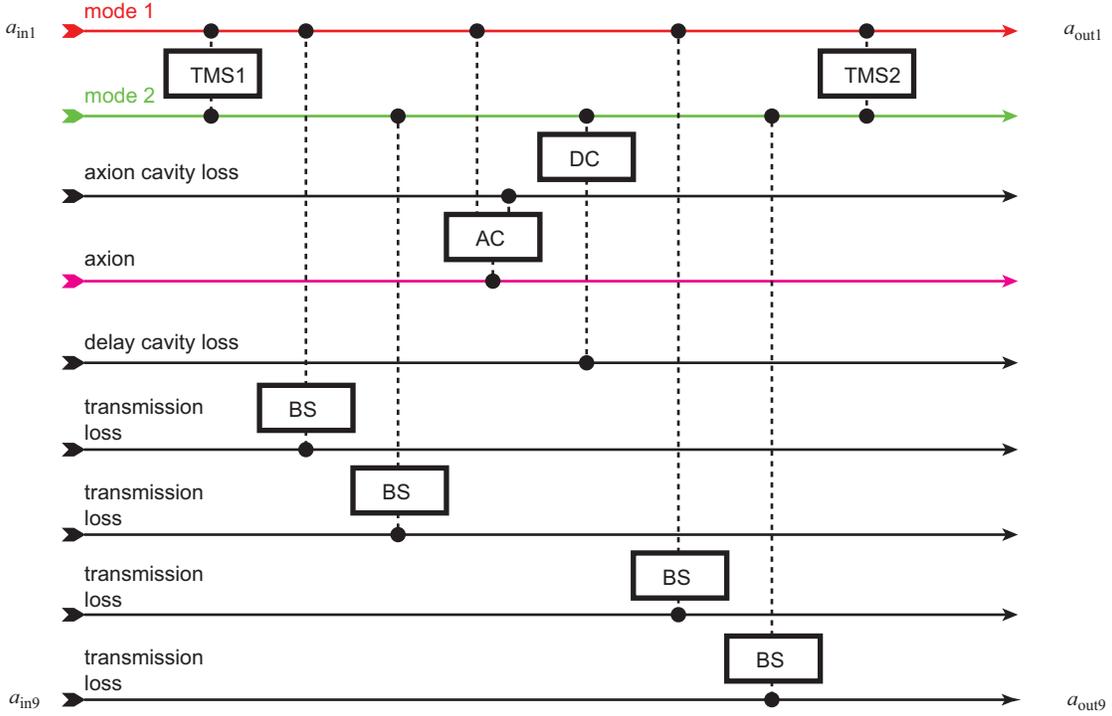}\\
  \caption{(Color online) Quantum optics schematic. The figure shows schematically the unitary scattering between the modes of the quantum optics model. Scattering occurs between modes (horizontal lines) joined (dots) by vertical lines. As indicated by the arrows, the modes propagate from left to right, which can be arranged experimentally by using circulator elements. The particular transformation is denoted by the boxed text acronym: TMS1 ( first two-mode squeezer), TMS2 ( second first two-mode squeezer), BS (beam splitter loss), AC (axion cavity), and DC (delay cavity). The first two modes are the target modes that the TMSs will entangle and disentangle. Mode 3 models the cavity loss, mode 4 the putative axion coupling, mode 5 the delay cavity loss and modes 6-9 the transmission losses.}\label{fig:EightModeModel}
\end{figure*}

The full quantum optics model for the two TMS receiver describes the scattering between nine field modes. We construct the $18\times18$ susceptibility matrix $\mathbf{\Xi}$ that performs the linear and unitary transformation $\vec{y}=\mathbf{\Xi} \vec{u}$, by cascading simpler transformations associated with the TMS elements, the beam splitters, and the cavities, as illustrated schematically in Fig.~\ref{fig:EightModeModel}. These simpler transformations are described in the following sections.

\subsubsection{Two-mode squeezer}
As described in section~\ref{sec:TwoModeSqueezer}, if the TMS couples modes $j$ and $k$, it transforms an input vector $\vec{v}_{\mathrm{in}}$ of the same form as $\vec{u}$ to an output vector $\vec{v}_{\mathrm{out}}$, as
\begin{eqnarray}\label{eq:TMSxform}
v_{\mathrm{out},j}&=&v_{\mathrm{in},j}\sqrt{G} + v_{\mathrm{in},19-k} \sqrt{G-1} \exp(i\phi)\\ \nonumber
v_{\mathrm{out},k}&=&v_{\mathrm{in},k}\sqrt{G} + v_{\mathrm{in},19-j}\sqrt{G-1} \exp(i\phi)\\ \nonumber
v_{\mathrm{out},19-j}&=&v_{\mathrm{in},19-j}\sqrt{G} + v_{\mathrm{in},k}\sqrt{G-1} \exp(-i\phi)\\ \nonumber
v_{\mathrm{out},19-k}&=&v_{\mathrm{in},19-k}\sqrt{G} + v_{\mathrm{in},j}\sqrt{G-1} \exp(-i\phi)\\ \nonumber
v_{\mathrm{out},l}&=&v_{\mathrm{in},l};\mathrm{if} \ l\ \mathrm{not\ in}\ \{j,k,19-j,19-k\}
\end{eqnarray}
where (for example) $v_{\mathrm{out},j}$ is the $j$-th element of the output vector. Note that the 3rd and 4th equations are just the Hermitian conjugate of the 1st and 2nd, and the last equation leaves unchanged modes which are neither the $j$-th nor the $k$-th. For the first TMS (TMS1), $j=1$, $k=2$, $G=G_1$, and $\phi=0$. For the second TMS (TMS2), $j=1$, $k=2$, $G=G_2$, and $\phi=\phi_{2}=\pi$. For the plots in Fig.~\ref{fig:SqueezBenA} and~\ref{fig:SqueezBen}, we choose $G_2=1000$.

\subsubsection{Beam-splitter loss}
Loss in transporting states between separate elements is modeled by beam splitter (BS) elements that couple modes 1 and 2 to unmeasured modes. If the BS couples modes $j$ and $k$, it performs the following transformation on vector elements $j$ and $k$
\begin{eqnarray} \label{eq:BSxform}
v_{\mathrm{out},j}&=&v_{\mathrm{in},j}\sqrt{\eta} + i v_{\mathrm{in},k}\sqrt{1-\eta}\\ \nonumber
v_{\mathrm{out},k}&=&v_{\mathrm{in},k}\sqrt{\eta} + i v_{\mathrm{in},j}\sqrt{1-\eta},
\end{eqnarray}
with the related transformation for elements $19-j$ and $19-k$ found by the Hermitian conjugate of the these equations, and with all other elements unchanged. For simplicity we assume that all beam splitter elements are described by a single power transmission value $\eta$.

\subsubsection{Axion cavity}

The axion cavity transformation has already been described in Eq.~\ref{eq:OutputAxionSuscept}. To incorporate $\chi_{jk}(\omega)$ into the model we need only associate the measurement port with mode 1, the loss port with mode 3, and the axion port with mode 4. The transformation to vector elements $j,k\in\{1,3,4\}$ is
\begin{equation}\label{eq:AxCavXform2}
v_{\mathrm{out},j} = \sum_{k}\chi_{jk}v_{\mathrm{in},k}(\omega),
\end{equation}
where the related transformation on indices 15, 16, and 18 can again be found from the Hermitian conjugate. All other elements are unchanged.

To complete the model of the axion cavity we estimate the value of $\ka$ and $\na$ to confirm that $\ka \ll \kl$ as we assumed in the noise analysis. These are only calculable within the various axion theories, which differ in their predictions for $g_{\gamma \gamma \mathcal{A}}$ by order unity factors \cite{Dine1981,KimPRL79}. In addition they depend directly on the axion mass. Nevertheless, a simple estimate suffices to see that $\ka$ is exceedingly small and our assumptions are valid. If the dark matter energy density $\rho_{E} \approx 0.45\ \mathrm{GeV/cm^3}$ is composed entirely of axions with rest mass $m_a c^2/h = 5$~GHz then the axion number density is $\rho_{N} \sim 3\times10^{13}/\mathrm{cm^3}$. The axion cavity will never deplete this source dark matter, a situation that we model by imagining a coherent field driving the axion-port of our cavity with amplitude $|\langle \hat{a}_{\mathrm{in,ax}}\rangle|=\sqrt{\ka \rho_{N} U_{\mathrm{vol}}}$, with $U_{\mathrm{vol}}$, the axion cavity volume. If a critically coupled axion cavity has been tuned into resonance with this axion field, we can evaluate $|\chi_{\mathrm{ma}}(\omega = 0)|^2$ to find a flux of axion-generated microwave photons emerging from the cavity at a rate $n_{\gamma} = (\ka^2 \rho_N U_{\mathrm{vol}})/\kl$. Various authors give experimentally parametrized expressions for the axion-derived microwave power emitted by an axion cavity. Using the characteristic parameters given in equation~14 of~\cite{BradleyRMP03} including $U_{\mathrm{vol}}=5\times10^5\ \mathrm{cm^3}$, but assuming $m_a c^2/h = 5$~GHz and $Q_L = \wc/(2 \kl) = 10^4$, we find $n_{\gamma} = 75$~photons/s. This value gives an estimate of $\ka/2\pi \sim 10^{-6}$~Hz, which is utterly negligible compared to $\kl/2 \pi = 2.5 \times 10^5$~Hz. Finally, we estimate $|\langle \hat{a}_{\mathrm{in,ax}}\rangle|^2 = \ka \rho_N U_{\mathrm{vol}} \sim 10^{12}$~axions/s. The axion field is believed to be spectrally broadened with a fractional linewidth of $10^{-6}$ or $(m_a c^2/h)/10^{6} = 5$~kHz \cite{SikiviePRD85}. The axion number-spectral-density would then be $\na \sim |\langle \hat{a}_{\mathrm{in,ax}}\rangle|^2/(5\ \mathrm{kHz}) \sim 2\times10^{10}$, an exceedingly weakly coupled but large amplitude field, thus justifying its treatment as a classical field.

\subsubsection{Delay cavity}

The delay cavity ensures that at each frequency $\omega$, mode 2 experiences the same phase shift and attenuation as mode 1. Except for the absence of (negligible) coupling to the axion field, the delay cavity should have the same susceptibility as the axion cavity. Thus the susceptibilty matrix for the delay cavity is
\begin{equation}\label{eq:DelCavXform}
\chi_{\mathrm{d},jk} = \frac{-\sqrt{\kappa_j}\sqrt{\kappa_k} + \left(\kappa_\mathrm{d}/2 + i \omega\right)\delta_{jk}}{\left(\kappa_\mathrm{d}/2 + i \omega\right)},
\end{equation}
with $\kappa_j$ and $\kappa_k \in \{\kappa_{\mathrm{dm}},\kappa_{\mathrm{dl}}\}$ and $\kappa_\mathrm{d} = \kappa_\mathrm{{dm}} + \kappa_\mathrm{{dl}}$. To match the axion cavity susceptibility we choose $\kappa_\mathrm{{dl}} = \kl$ and $\kappa_\mathrm{{dm}} = \kc$. Mode 5 is associated with the loss port of the delay cavity and mode 2 with the measurement port. The transformation to vector elements with $j,k\in\{2,5\}$ is
\begin{equation}\label{eq:DelCavXform2}
v_{\mathrm{out},j} = \sum_{k}\chi_{\mathrm{d},jk}v_{\mathrm{in},k}(\omega),
\end{equation}
where the related transformation on indices 14 and 17 can again be found from the Hermitian conjugate. All other elements are unchanged.

Because the delay cavity has no coupling to the axion field, it need not reside in a magnetic field; thus, it could have negligible loss compared to the axion cavity.  Giving the cavities the same susceptibility ensures that at all frequencies the two modes experience the same delay and same attenuation passing through their respective cavities. This balance is crucial to the operation of the two-mode squeezing concept. Although the deliberate introduction of loss to a quantum measurement is surprising, the benefit of loss is apparent when considering the limit of a critically coupled axion cavity at resonance. Without loss in the delay cavity, it will reflect one half of the two-mode squeezed state while the axion cavity perfectly absorbs the other half. In this case, squeezing is not only unhelpful; it is detrimental because the first TMS simply adds noise to the measurement performed by the second TMS.  

In contrast, when the transmission loss is low $(1-\eta \ll 1)$, one can operate in the limit of a very overcoupled axion cavity. In this limit, balancing attenuation is less important as the two modes will suffer very little attenuation anyway and a lossless delay cavity with $\kappa_\mathrm{{dl}}=0$ and $\kappa_\mathrm{{dm}}=\kappa(1-(\kl/\kc))$ yields a greater search rate. (This choice matches the delay of the two modes exactly). For very low transmission loss $(\eta>0.9)$, a lossless delay cavity yields a search rate enhancement more than 5\% greater than a lossy delay cavity. For transmission loss $(\eta \leq 0.9)$, numerically optimizing the search rate over the parameters $(\kc,\ \kappa_\mathrm{{dm}},\ \kappa_\mathrm{{dl}},\ G_1)$, yields less than a 5\% benefit over choosing $\kappa_\mathrm{{dm}}=\kc$ and $\kappa_\mathrm{{dl}}=\kl$.  Here we do not consider the possibility of replacing the TMS elements with objects that would preform the most general unitary, two-mode transformations although it seems possible that such a generalization could overcome the unbalanced attenuation of the two modes without adding loss to the delay cavity.

\bibliography{AxionSearchSPDandSqeezingBib_kl}

\end{document}